\newif\ifcomment
\newif\ifdraft
\newif\iflatexdiff
\def\dvers{v1.0}
\def\dtitle{Direct photon production at low transverse momentum\\ in proton--proton collisions at $\mathbf{\sqrt{s}=2.76}$ and 8 TeV}
\def\stitle{Direct photon production at low $\pT$ in pp at $\sqrt{s}=2.76$ and 8 TeV} 
\documentclass[ALICE,manyauthors]{cernphprep}
\usepackage[comma,square,numbers,sort&compress]{natbib}
\usepackage{hyperref}
\usepackage{graphicx}  
\usepackage{dcolumn}   
\usepackage{bm}        
\usepackage{amssymb}   
\usepackage{amsfonts}
\usepackage{graphics}
\usepackage{grffile}   
\usepackage{epsfig}
\usepackage[loose]{units}
\usepackage[usenames,dvipsnames]{xcolor}
\usepackage[normalem]{ulem} 
\usepackage{color}
\usepackage[utf8]{inputenc}
\usepackage[T1]{fontenc}
\usepackage{subfigure}
\usepackage{mathrsfs}
\usepackage{multirow}
\iflatexdiff
\RequirePackage{color}\definecolor{RED}{rgb}{1,0,0}\definecolor{BLUE}{rgb}{0,0,1}

\fi

\newcommand{\lzt}  	   {\ensuremath{\sigma^{2}_{\rm long}}}
\newcommand{\Vz}  	   {\ensuremath{V^{0}}}

\newcommand{\kzs}          {K$^{0}_{\rm S}$}
\newcommand{\kzl}          {K$^{0}_{\rm L}$}

\newcommand{\pp}           {pp}

\newcommand{\pPb}          {\mbox{p--Pb}}

\newcommand{\PCM}          {\mbox{PCM}}

\newcommand{\EMC}          {\mbox{EMC}}

\newcommand{\PCMEMC}       {\mbox{PCM-EMC}}

\newcommand{\gevc}         {GeV/$c$}
\newcommand{\Rg}           {\ensuremath{R_{\gamma}}}
\newcommand{\Yd}           {\ensuremath{Y_{\gamma_{\rm dir}}}}
\newcommand{\Ygi}          {\ensuremath{Y_{\gamma_{\rm \,incl}}}}
\newcommand{\Ygd}          {\ensuremath{Y_{\gamma_{\rm decay}}}}
\newcommand{\Yp}           {\ensuremath{Y_{\pi^0}}}
\newcommand{\s}            {\ensuremath{\sqrt{s}}}

\newcommand{\pt}           {\ensuremath{p_{\mathrm{T}}}}
\newcommand{\pT}           {\pt}
\newcommand{\mT}           {\ensuremath{m_{\mathrm{T}}}}
\newcommand{\GeVc}         {\ensuremath{\mbox{GeV}/c}}

\newcommand{\dedx}         {d$E$/d$x$}

\newcommand{\mt}           {\ensuremath{m_{\mathrm{T}}}}
\newcommand{\snn}          {\ensuremath{\sqrt{s_{\mathrm{NN}}}}}

\newcommand{\Fig}[1]       {Fig.~\ref{#1}}

\newcommand{\Sect}[1]      {Sec.~\ref{#1}}
\newcommand{\Section}[1]   {Section~\ref{#1}}

\newcommand{\Eq}[1]        {Eq.~\ref{#1}}

\newcommand{\Tab}[1]       {Tab.~\ref{#1}}

\newcommand{\Ref}[1]       {Ref.~\cite{#1}}
\newcommand{\Refs}[1]      {Refs.~\cite{#1}}

\newcommand{\com}[1]       {}
\iflatexdiff

\renewcommand{\xout}[1]    {\textcolor{red}{\sout{#1}}}
 
\newcommand{\old}[1]       {{\textcolor{red}{\sout{#1}}}}

\else

\renewcommand{\xout}[1]    {}
\newcommand{\old}[1]       {\relax}

\fi

\graphicspath{{./img/}}
\ifdraft
\usepackage{lineno}
\linenumbers
\modulolinenumbers[1]
\usepackage{fancyhdr}
\fi
\begin{document}
\begin{titlepage}
\PHyear{2018}
\PHnumber{045}    
\PHdate{18 March}      
\title{\dtitle}
\ShortTitle{\stitle}
\Collaboration{ALICE Collaboration%
         \thanks{See Appendix~\ref{app:collab} for the list of collaboration members}}
\ShortAuthor{ALICE Collaboration} 
\begin{center}
\ifdraft
\today\\ \color{red}DRAFT \dvers\ \hspace{0.3cm} \$Revision: 4288 $\color{white}:$\$\color{black}\vspace{0.3cm}
\else
\fi
\end{center}
\begin{abstract}
  Measurements of inclusive and direct photon production at mid-rapidity in pp collisions at $\sqrt{s}=2.76$ and 8 TeV are presented by the ALICE experiment at the LHC.
  The results are reported in transverse momentum ranges of $0.4<\pT<10$ \GeVc\ and $0.3<\pT<16$ \GeVc, respectively.
  Photons are detected with the electromagnetic calorimeter~(EMCal) and via reconstruction of e$^+$e$^-$ pairs from conversions in the ALICE detector material using the central tracking system.
  For the final measurement of the inclusive photon spectra the results are combined in the overlapping $\pT$ interval of both methods. 
  Direct photon spectra, or their upper limits at 90\% C.L.\ are extracted using the direct photon excess ratio $\Rg$, which quantifies the ratio of inclusive photons over decay photons generated with a decay-photon simulation.
  An additional hybrid method, combining photons reconstructed from conversions with those identified in the EMCal, is used for the combination of the direct photon excess ratio $\Rg$, as well as the extraction of direct photon spectra or their upper limits.
  While no significant signal of direct photons is seen over the full $\pT$ range, $\Rg$ for $\pT>7$ \GeVc\ is at least one $\sigma$ above unity and consistent with expectations from next-to-leading order pQCD calculations.
\end{abstract}
\end{titlepage}
\newpage
\setcounter{page}{2}
\section{Introduction}
\label{sec:intro}
Major experimental efforts are undertaken at the Relativistic Heavy Ion Collider (RHIC)~\cite{Back:2004je,Arsene:2004fa,Adcox:2004mh,Adams:2005dq} and the Large Hadron Collider (LHC)~\cite{Aamodt:2010pb,Aamodt:2010cz,Chatrchyan:2011sx,Aad:2010bu,Abelev:2012hxa,ATLAS:2011ah,Aamodt:2010pa,Chatrchyan:2011pb,ATLAS:2011ag} to study the conditions for the creation and the properties of the quark-gluon plasma (QGP), a deconfined partonic state predicted by the theory of strong interaction, Quantum ChromoDynamics~(QCD)~\cite{Bazavov:2011nk,Borsanyi:2013bia}. 
Direct photons, which are defined as all photons that are produced directly in scattering processes and therefore do not originate from hadronic decays, are a powerful tool for exploring the QGP.
They are produced during all stages of the collision and are basically unaffected by final state interactions as they only participate in electromagnetic interactions~\cite{Thoma:1994fd}.
Hence, they are sensitive to the early stages of the collision's evolution.
Since a variety of QGP signatures are also present in high multiplicity \pPb\ or \pp\ collision at the LHC~\cite{Loizides:2016tew}, it is interesting to study if a direct photon signal at low $\pT$ can be observed already in minimum bias pp collisions, as predicted for $\sqrt{s}=7$ TeV~\cite{Liu:2011dk}.
Experimentally, however, the main challenge for direct photon measurements is to distinguish them from the large background of decay photons.

Depending on their production mechanism, direct photons are usually classified into two main categories: prompt and thermal photons.
Prompt photons carry information about parton distributions in nuclei~\cite{PhysRevD.26.3284,PhysRevD.40.3128} as they are produced in hard scatterings of incoming partons, such as Compton scattering $q+g\rightarrow q+\gamma$ or annihilation $q+\overline{q}\rightarrow g+\gamma$, as well as bremsstrahlung emission from quarks which undergo a hard scattering~\cite{PhysRevLett.90.132301,PhysRevC.77.024909,Arleo:2011gc}.
These processes are described by perturbative QCD~(pQCD) in leading and next-to-leading order, which are dominant at LHC energies.
One of the purposes of direct photon measurements is to improve the accuracy of such calculations in various collision systems.
At RHIC at center of mass energies per nucleon-nucleon pair of $\snn=0.2$~TeV and at the LHC at $\snn=2.76$~TeV, direct photons with transverse momenta~($\pT$) above about $3$ and $15$~\GeVc, respectively, were found to be dominated by prompt photons and to follow a power law spectral shape in small systems (pp, pA, dA)~\cite{Abelev:2009hx,Adare:2012vn,Aad:2010sp,Chatrchyan:2012vq} as well as in heavy-ion collisions~\cite{Chatrchyan:2012vq,Afanasiev:2012dg,Adare:2014fwh,Aad:2015lcb}, as described by pQCD.

In heavy-ion collisions, additionally, thermal photons are expected to be radiated off the locally thermalized, hot QGP and hadronic matter, which provide information about the temperature, collective expansion, as well as the space-time evolution of the medium~\cite{Gale:2009gc}, and are expected to dominate the direct photon spectrum at low transverse momenta~($\pT\lesssim3$~\GeVc)~\cite{PhysRevD.44.2774,Turbide:2003si}.
Further direct photon production mechanisms, like interactions of hard scattered partons with dense partonic matter~(``jet-photon conversion'')~\cite{Fries:2002kt,Turbide:2007mi}, as well as production of photons from non-equilibrated phases~\cite{Vogelsang:1997cq}, may also play a role in the low and intermediate \pT\ region from $3$ to ${\sim}10$ \GeVc.

In the following, we present first results from the measurement of direct photon production at mid-rapidity in $0.4<\pT<10$ \GeVc\ and $0.3<\pT<16$ \GeVc\ in pp collisions at $\sqrt{s}=2.76$ and 8 TeV, respectively.
These data, which are the first direct photon data below 15~\GeVc\ in pp collisions at the LHC, enable pQCD calculations to be tested in this low $\pT$ regime.
Furthermore, they provide an important baseline for the interpretation of initial- and final-state effects observed in direct photon data from heavy-ion collisions~\cite{Adare:2011zr,Adam:2015lda}, because event generators and perturbative calculations are generally not reliable at $\pT\lesssim3$.

The direct photon yield is extracted by comparing the measured inclusive photon spectrum to the spectrum of photons from hadron decays via a double ratio, obtained from the so-called ``direct photon excess ratio''~$\Rg$~\cite{Aggarwal:2000ps,Aggarwal:2000th}.
The double ratio, defined on the level of fully corrected quantities, can be written as
\begin{equation}
 \Rg=\frac{\Ygi}{\Ygd} \approx \left(\frac{\Ygi}{\Yp}\right)_\text{meas}/\left(\frac{\Ygd}{\Yp}\right)_\text{sim}\,,
 \label{eq:Rg}
\end{equation}
where the numerator denotes the measured inclusive photon yield, $\Ygi$, divided by the measured neutral pion yield, $\Yp$, and the denominator is constructed in the same way, but with the photon yield obtained by a decay-photon simulation and a parameterization of the neutral pion yield. 
With $\Rg$, the direct photon yield can then be obtained from the inclusive photon yield as
\begin{equation}
 \Yd=\Ygi-\Ygd=\left(1-\frac{1}{\Rg}\right) \, \Ygi\,.
 \label{eq:gd}
\end{equation}
The yields in \Eq{eq:Rg} and \Eq{eq:gd} are implicitly defined at midrapidty as a function of \pT\ of the corresponding particle.
The cross sections can be obtained by replacing the inclusive photon yield with the inclusive photon cross section in \Eq{eq:gd}.  
The advantage of using $\Rg$~(rather than trying to directly quantify the difference of inclusive and decay photons) is the partial or full cancellation of several systematic uncertainties in the double ratio.
The photon reconstruction is performed independently using either conversions in the inner detector material reconstructed with the central tracking system and for the first time with the Electromagnetic Calorimeter~(EMCal).
Combined inclusive and direct photon spectra are determined based on the individual inclusive photon spectra and direct photon excess ratios.
The direct photon spectra, or respectively their upper limits at 90\% C.L., are finally compared to next-to-leading order pQCD calculations.

The paper is structured as follows:
\Section{sec:cocktail} describes the photon-decay simulation at generator level, usually known as ``cocktail simulation''.
\Section{sec:aliceDet} describes the relevant ALICE detectors for the photon and neutral meson measurements, the data taking conditions, and the event selection. 
\Section{sec:photonreco} describes the data analysis with emphasis on the photon reconstruction via the Photon Conversion Method~(PCM) and using the EMCal.
The systematic uncertainties are summarized in \Section{sec:systematics}, whereas \Section{sec:results} presents the results.
\Section{sec:conclusion} concludes with a short summary.

\begin{figure}[t]
        \center
        \includegraphics[height=.53\textwidth]{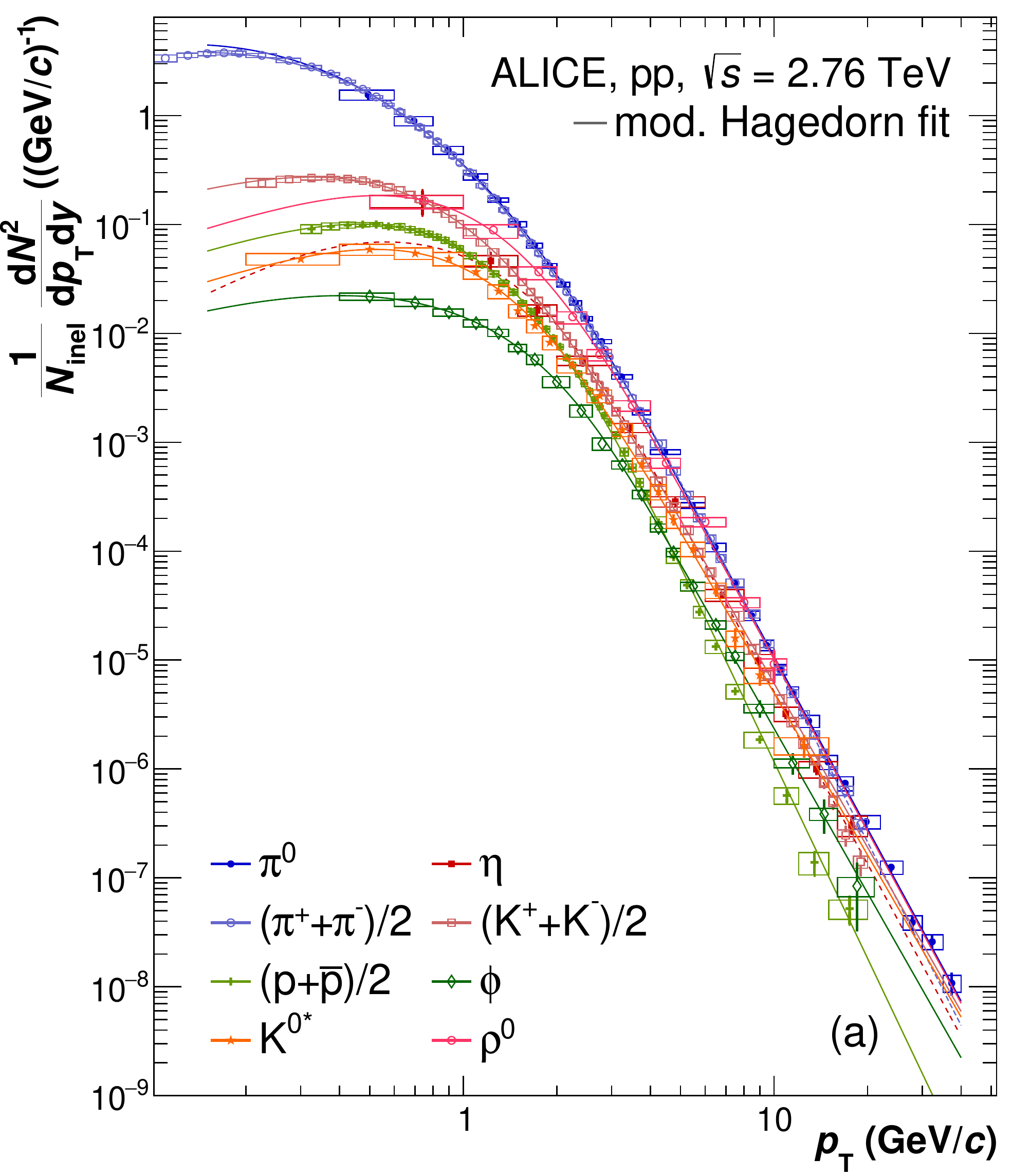}
        \hspace{0.5cm}
        \includegraphics[height=.53\textwidth]{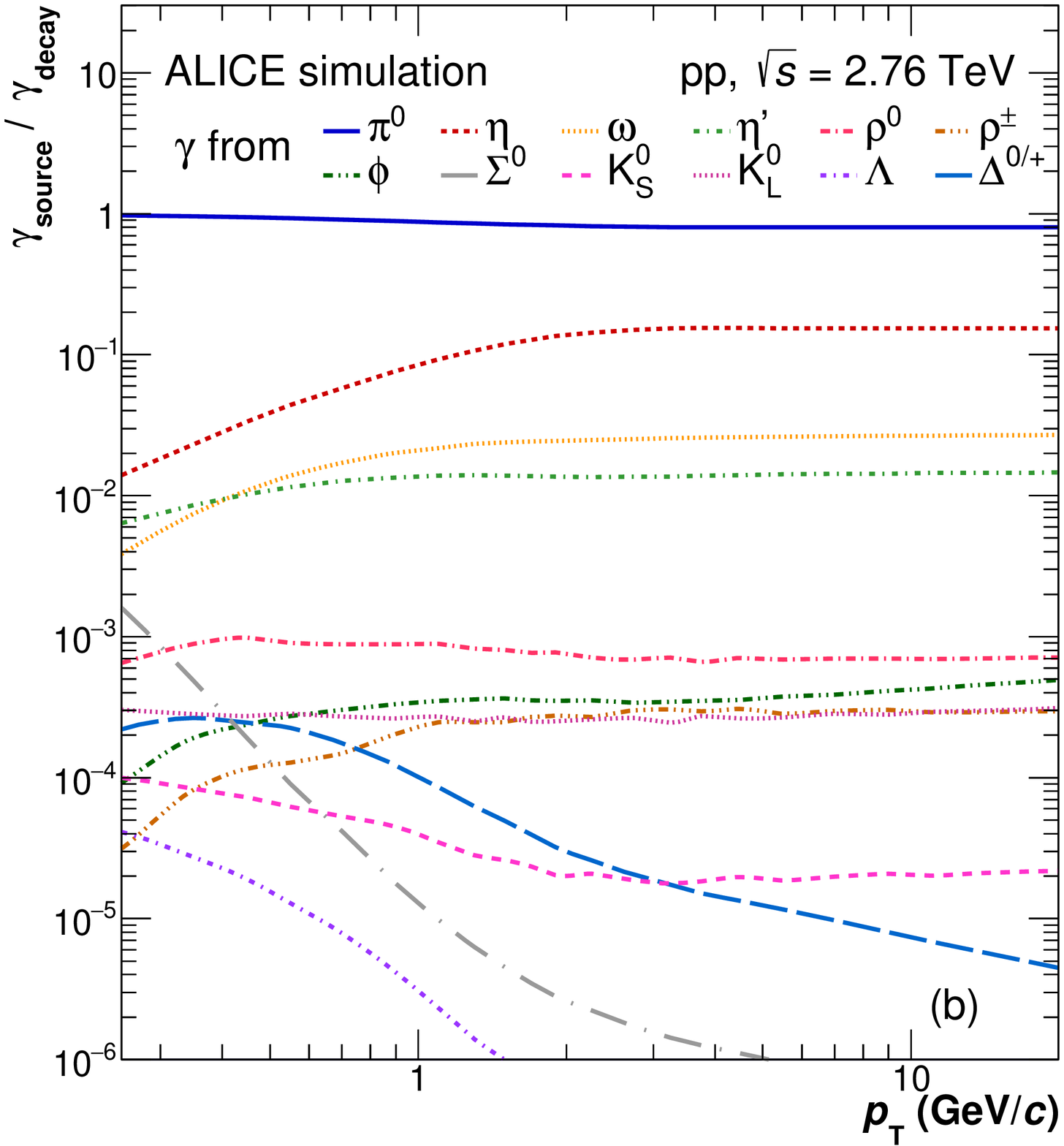}
        \caption{\textit{(a)} Measured identified particle yields per inelastic event in pp collisions at $\sqrt{s}=2.76$ TeV~\cite{Acharya:2017hyu,Abelev:2014laa,Adam:2017zbf,Acharya:2018qnp}
                         including their modified Hagedorn parametrization used as input for the cocktail simulation. Statistical uncertainties are shown with vertical lines and systematic 
                         uncertainties with boxes. See the text for references to the data.
                 \textit{(b)} Ratio of primary decay photons from different sources to all primary decay photons in the decay photon simulation for pp collisions at $\sqrt{s}=2.76$ TeV. From top to bottom at high \pT\, the different sources are $\pi^0$, $\eta$, $\omega$, $\eta'$, $\rho^0$, $\phi$, \kzl, $\rho^\pm$, \kzs, $\Delta^{0/+}$, $\Sigma^0$ and $\Lambda$. }
        \label{fig:inputspectraandcocktailgammas}
\end{figure}

\section{Generator-level decay-photon simulation}
\label{sec:cocktail}
The decay photon spectra are obtained by a particle decay simulation, also called ``cocktail simulation'', needed for the secondary decay photon correction as well as for the calculation of $\Rg$.
The decay simulation is based on the PYTHIA 6.4 particle decayer~\cite{Sjostrand:2006za} with random generation of mother particles uniform in azimuth and $\pT$.
Parametrizations of the transverse momentum spectra of the mother particles measured by ALICE are used as weights in order to obtain the correct abundances.

For the $\sqrt{s}=2.76$ TeV cocktail, measured \pT\ differential yields per inelastic event of $\pi^0$~\cite{Acharya:2017hyu}, K$^\pm$, p~\cite{Abelev:2014laa}, $\phi$~\cite{Adam:2017zbf} and $\rho^0$~\cite{Acharya:2018qnp} as well as the $\eta/\pi^0$~\cite{Acharya:2017hyu} ratio are parametrized as inputs, and are shown in the left panel of \Fig{fig:inputspectraandcocktailgammas}.
Neutral kaons, which constitute an important background for secondary decay photons, are approximated by the average of the charged kaon yields. 
The particle decay simulation for $\sqrt{s}=8$ TeV uses measured \pT\ differential yields from $\pi^0$ and $\eta$~\cite{Acharya:2017tlv} as input.
Furthermore, \pT\ differential yields for K$^\pm$, $\phi$, and p are extrapolated using the measured spectra at $\sqrt{s}=2.76$ and 7 TeV~\cite{Adam:2015qaa,Abelev:2012hy,ALICE:2017jyt} as inputs. 
The extrapolation is done on a bin-by-bin basis in \pT\ assuming a power-law evolution of the particle yields with increasing center-of-mass energy.
For the parametrization, the \pT\ differential particle yields are fitted with a modified Hagedorn function~\cite{Hagedorn1983,Altenkamper:2017qot} whose functional form is given by 
\begin{eqnarray}
	\frac{\text{d}^2N}{\text{d}y\text{d}\pt} = \pt \cdot A \cdot \left( \exp{\left( a\pt + b\pt^2 \right)} + \frac{\pt}{p_0} \right)^{-n}.
	\label{eq:modifiedHagedorn}
\end{eqnarray}
In order to obtain a stable parametrization for the $\eta$ particle yields up to high \pT, the $\eta/\pi^0$ ratios at $\sqrt{s}=2.76$ and 8 TeV are fitted with an empirical function that describes contributions from soft and hard processes~\cite{Altenkamper:2017qot}, given as
\begin{eqnarray} 
	\frac{\eta}{\pi^0}(\pt) = \frac{ A \cdot \exp \left( \frac{ \beta \pt - \mt^{\eta} }{ T \sqrt{1 - \beta^2} } \right) + N  \cdot B \cdot \left( 1 + \left( \frac{\pt}{p_0} \right)^2 \right)^{-n} }{ \exp \left( \frac{ \beta \pt - \mt^{\pi^0} }{ T \sqrt{1 - \beta^2} } \right) + B \cdot \left( 1 + \left( \frac{\pt}{p_0} \right)^2 \right)^{-n} },
	\label{eq:softHard}
\end{eqnarray}
with a relative normalization factor $B$ between the soft and hard part of the parametrization and the constant ratio value $N$ between the two particle species that is approached at high $\pT$.
All spectra are described by their parametrization within a maximum of 10\% deviation over the full transverse momentum range.
For particles that are neither measured nor extrapolated, the parametrization is obtained via transverse mass scaling, $\mt = \sqrt{\pT^2 + m_0^2}$, with the neutral pion as basis ($B$) for mesons and the proton as basis for baryons.
The $\mt$ scaling factors, $C_{\mt}^X=(\mathrm{d}N_{X}/\mathrm{d}\mt)/(\mathrm{d}N_{B}/\mathrm{d}\mt)$, for each particle $X$ are derived from the respective spectra in PYTHIA.
These particles are $\eta'$ ($C_{\mt}=0.4$), $\Lambda$ ($C_{\mt}=1.0$), $\Sigma^0$ ($C_{\mt}=0.49$), $\Delta^{0,+}$ ($C_{\mt}=1.0$) and $\omega$ ($C_{\mt}=0.85$) at $\sqrt{s}=2.76$ TeV and additionally $\rho^0$ ($C_{\mt}=1.0$) at $\sqrt{s}=8$ TeV.
The limitations of transverse mass scaling~\cite{Altenkamper:2017qot} at low \pT\ can be neglected for this measurement as the transverse mass scaled particles contribute only a tiny fraction to the total decay photon yield as seen in \Fig{fig:inputspectraandcocktailgammas}.

For the particle decay simulation, particles are generated uniformly in the transverse momentum range of $0\leq\pT\leq50$ \GeVc, for the rapidity range of $|y|<1.0$ as well as the full azimuth of $0<\phi<2\pi$.
For each particle, the full decay chain is simulated which allows the decay simulation to be used for the secondary photon correction for photons produced following weak decays of primary hadrons as well as the extraction of the decay photon spectrum.
After generation, only decay photons are kept, which fulfill $|y|<0.9$, to match the hadron rapidity range that is used in the measurements.
Furthermore, the mother particles as well as all decay products are weighted with the parametrization of the mother particle.
The contribution of each individual decay photon source to all decay photons of the cocktail simulation is shown in the right panel of \Fig{fig:inputspectraandcocktailgammas}.
Decay photons originating from $\pi^0$ decays are the dominant contribution with $\sim 86$\% of total decay photons at high \pT.
Contributions from the $\eta$ meson decay photons represent $\sim 10\%$ whereas decay photons from $\omega$ and $\eta'$ mesons contribute below 3\% and 1.5\%, respectively.
All other, remaining sources are basically negligible as shown in \Fig{fig:inputspectraandcocktailgammas}.

\section{Experimental setup and data taking conditions}
\label{sec:aliceDet}
Two different methods using independent detector systems of ALICE~\cite{Aamodt:2008zz} are employed to measure photons in this analysis.
In the first method, PCM, photons are reconstructed from e$^+$e$^-$ pairs, which are created by photon conversions in the inner detector material.
The inner  material includes the full active and passive material of the beampipe, the Inner Tracking System (ITS) as well as the inner field cage vessel of the Time Projection Chamber~(TPC) and part of the TPC gas. 
The main tracking systems in ALICE at mid-rapidity, the ITS and the TPC, are used for the reconstruction of these electron-positron pairs, which originate from secondary vertices~(V$^0$).
In the second photon reconstruction method, \EMC, the energy deposit in the EMCal is used to measure photons.
With PCM, a high momentum resolution at low \pT\ is achieved, but the method is limited by statistics at high transverse momenta.
The \EMC\ method benefits from large statistics up to high \pT\ but has a decreasing resolution towards low \pT.
The necessary detector systems are described in the following with emphasis on the detector configurations in both pp data taking periods of $\sqrt{s}=2.76$ TeV in 2011 and $\sqrt{s}=8$ TeV in 2012.

The ITS~\cite{Aamodt:2010aa} consists of three sub-detectors each with two layers to measure the trajectories of charged particles and to reconstruct primary~\cite{ALICE-PUBLIC-2017-005} and secondary vertices~\cite{Alessandro:2006yt}.
The two innermost layers are the Silicon Pixel Detectors (SPD) positioned at radial distances of $3.9$~cm and $7.6$~cm relative to the beam line, followed by two layers of Silicon Drift Detectors (SDD) at $15.0$~cm and $23.9$~cm, and completed by
two layers of Silicon Strip Detectors (SSD) at $38$~cm and $43$~cm.
The two layers of SPD cover pseudorapidity ranges of $|\eta|<2$ and $|\eta|<1.4$, respectively.
The SDD and SSD cover $|\eta|<0.9$ and $|\eta|<1.0$, accordingly.

The TPC~\cite{Alme:2010ke} is a large (90~${\rm m}^{3})$ cylindrical drift detector filled with ${\rm Ne}$-${\rm CO_{2}}$ $(90$\%-$10\%)$ gas mixture.
It covers a pseudorapidity range of $|\eta|<0.9$ over the full azimuth, providing up to 159 reconstructed space points per track.
A magnetic field of $B=0.5$~T is generated by a large solenoidal magnet surrounding the central barrel detectors.
Charged tracks originating from the primary vertex can be reconstructed down to $\pT\approx 100$~MeV/$c$ and charged secondaries down to $\pT\approx 50$~MeV/$c$ with a tracking efficiency of $\approx80\%$ for tracks with $\pT> 1$~GeV/$c$~\cite{Abelev:2014ffa}.
In addition, the TPC provides particle identification via the measurement of energy loss \dedx\ with a resolution of ${\approx}5\%$.
The ITS and TPC are complemented by the Transition Radiation Detector~(TRD)~\cite{Acharya:2017lco} and a large Time-Of-Flight~(TOF)~\cite{Adam:2016ilk} detector.

The EMCal detector~\cite{Cortese:2008zza} is an electromagnetic sampling calorimeter covering $\Delta\phi=100^\circ$ in azimuth and $|\eta|<0.7$ in pseudorapidity, located at a radial distance of 4.28 m from the nominal collision vertex.
During the data taking periods in 2011 and 2012, it consisted of a total of 11,520 active elements, or cells, each of which comprise 77 alternating layers of lead and plastic scintillator providing a radiation length of $20.1\,X_{0}$.
Attached perpendicular to the face of each cell are wavelength shifting fibers that collect the scintillation light in each layer.
Avalanche Photo Diodes (APDs) with an active area of $5\times 5$~mm$^{2}$ are connected to the fibres to detect the generated scintillation light.
The size of each cell is $\Delta \eta \times \Delta \phi = 0.0143 \times 0.0143$ rad ($\approx6.0\times 6.0$~cm$^{2}$), corresponding to approximately twice the Molière radius. 
The EMCal consists of ten supermodules, where each supermodule is composed of $12 \times 24$ modules, consisting of $2 \times 2$ cells apiece.
It has an intrinsic energy resolution of $\sigma_{E}/E = 4.8\%/E \oplus 11.3\%/\sqrt{E} \oplus 1.7\%$ where the energy $E$ is given in units of GeV~\cite{Abeysekara:2010ze}.
The energy calibration of the detector is performed by measuring, in each cell, the reconstructed $\pi^0$ mass in the two-photon invariant mass distribution with one photon associated with the given cell. 
An estimated calibration level of better than 3\% is achieved with this method, which adds up quadratically to the constant term of the energy resolution.
Between 2011 and 2012 an additional TRD\com{Transition Radiation Detector~\cite{Acharya:2017lco}} module in front of EMCal was installed which results in a slightly different outer material budget between the $2.76$ and $8$~TeV data sets.
The material budget differences due to the TRD will be studied in detail for the estimation of the associated systematic uncertainties.

As trigger for minimum bias pp collisions and to reduce beam-induced background and pileup events the V0 detector~\cite{Abbas:2013taa} is used. 
It consists of two scintillator arrays~(V0A and V0C) covering $2.8<\eta<5.1$ and $-3.7<\eta<-1.7$. 
The probability of collision pile-up per triggered event was below 2.5\% and below 1\% at $\sqrt{s}=2.76$ and 8 TeV, respectively. 
Background events from beam-gas interactions or detector noise are rejected based on the timing information from V0A and V0C~\cite{Abelev:2014ffa}.
Events containing more than one pp collision within a single bunch crossing are rejected based on the information reconstructed in the SPD. 
In these events, either multiple primary vertices could be reconstructed within the acceptance~\cite{Abelev:2014ffa} or an excess of SPD clusters with respect to the number of SPD tracklets could be observed.
In addition, the primary vertex is required to be reconstructed within $|z|<10$ cm from the nominal interaction point. 
In 2011, the minimum bias trigger condition required a hit in either the SPD, the V0A or the V0C~(MB$_{\rm OR}$ condition), whereas in 2012 a hit in the V0A and the V0C~(MB$_{\rm AND}$ condition) was required.
The latter was necessary due to the higher beam intensities in 2012. 
The corresponding cross section for the minimum bias triggers are obtained from van der Meer scans~\cite{vanderMeer:1968zz} yielding $\sigma_{\rm MB_{OR}}=55.4\pm3.9_\text{stat+syst}$~mb~\cite{Abelev:2012sea} and $\sigma_{\rm MB_{AND}}=55.8\pm1.2_\text{stat}\pm1.5_\text{syst}$~mb~\cite{ALICE-PUBLIC-2017-002} for the data taking campaigns at $\sqrt{s}=2.76$ TeV and $\sqrt{s}=8$ TeV, respectively.
For the conversion-based measurements, integrated luminosities of $\mathscr{L_{\rm int}}=0.96\pm0.07_\text{norm}$ nb$^{-1}$ at $\sqrt{s}=2.76$ TeV and $\mathscr{L_{\rm int}}=2.17\pm0.06_\text{norm}$ nb$^{-1}$ at $\sqrt{s}=8$ TeV are analyzed.
The calorimeter-based measurements sample 50\% and, respectively, 10.6\% smaller integrated luminosities, since the EMCal was not always active during data taking.

Further information about the performance of these and other detector systems can be found in \Ref{Abelev:2014ffa}.

\section{Photon reconstruction}
\label{sec:photonreco}
Inclusive photons are reconstructed in two ways; either using photon conversions\com{ in the inner detector material} between $0.4$~($0.3$) and $8$~($16$)~\gevc\ or using the EMCal between $1.5$ and $10$~($16$)~\gevc\ for $\s=2.76$~($8$)~TeV.
Photons convert within the inner detector material of ALICE with a probability of about $8.9\%$, and are reconstructed with the PCM method as follows:
(i) tracking of charged particles and secondary vertex finding~\cite{Alessandro:2006yt};
(ii) particle identification and (iii) photon candidate reconstruction and subsequent selection.
The secondary vertices used in this analysis are obtained during data reconstruction by employing the full tracking capabilities of ITS and TPC.
For the daughter tracks, a minimum of 60\% of the maximum possible findable TPC clusters, that a particle track can create in the TPC along its path, and a minimum track \pT\ of 50 MeV/$c$ are required.
The contamination from Dalitz decays is reduced by rejecting conversion candidates with reconstructed vertices with a radial distance of less than $5$~cm with respect to the nominal center of the detector.
Furthermore, only secondary tracks and vertices with $|\eta|<0.9$ are accepted.
In addition, we restrict the geometrical $\eta$ distribution of the ${\rm V^{0}}$s in order to remove photon candidates that would otherwise appear outside the angular dimensions of the detector. 
To do so, the condition $R_{\rm conv}>|Z_{\rm conv}| S_{\rm ZR}-7$~cm is applied with $S_{\rm ZR}=\tan\left(2\arctan(\exp(-\eta_{\rm max}))\right)\approx0.974$ for $\eta_{\rm max}=0.9$, where $R_\text{conv}$ and $Z_{\rm conv}$ denote the radial and longitudinal coordinate of the conversion point, respectively.
The coordinates $R_\text{conv}$ and $Z_\text{conv}$ are determined with respect to the center of the detector and are set to $R_\text{conv}<180$~cm and $|Z_\text{conv}|<240$~cm to ensure a high quality secondary track reconstruction inside the TPC.

Electrons and positrons are identified via their energy deposit in the TPC, \dedx, by employing the difference of the measured \dedx\ to the expected value for electrons and positrons~\cite{Abelev:2014ffa}.
For the measurement at $\sqrt{s}=2.76$ TeV, the \dedx\ of the charged tracks is required to be within $-4<\text{n}_{\sigma_{\rm e}}<5$ of the expected electron/positron energy loss, where $\text{n}_{\sigma_{\rm e}}=($d$E/$d$x-\langle $d$E/$d$x \rangle_{\rm e})/\sigma_{\rm e}$ is \pT-dependent with the average energy loss of the electron/positron, $\langle $d$E/$d$x \rangle_{\rm e}$, and the Gaussian width of the fit to the measured \dedx\ distribution, $\sigma_{\rm e}$.
This condition is tightened for the measurement at $\sqrt{s}=8$ TeV to $-3<\text{n}_{\sigma_{\rm e}}<5$.
To reduce the contamination from pions, an additional selection based on the separation from the charged pion energy loss hypothesis is required in $\text{n}_{\sigma_\pi}$.
A rejection of tracks with energy losses closer to the pion line than $|\text{n}_{\sigma_\pi}|<1$ is applied up to a \pT\ of 3.5 \GeVc.
In the $\sqrt{s}=2.76$ TeV analysis, this rejection is continued above $\pT>3.5$~\GeVc\ with an $|\text{n}_{\sigma_\pi}|<0.5$ in order reduce the contamination even further in the momentum region where the two d$E$/d$x$ bands of the pion and electron merge.

Further contamination from non-photonic V$^0$ candidates is suppressed by a triangular two-dimensional selection range of $|\Psi_{\text{pair}}|<\Psi_{\text{pair,max}}(1-\chi_{\text{red}}^2/\chi_{\text{red,max}}^2)$ with $\chi_{\text{red,max}}^2=30$ and $\Psi_{\text{pair,max}}=0.1$~rad.
As explained in \Ref{Acharya:2017hyu}, this selection is based on the reduced $\chi^2$ of the Kalman-Filter hypothesis~\cite{Fruhwirth:1987fm,KalmanFilter} for the e$^+$e$^-$ pair and on the angle $\Psi_{\text{pair}}$ between the plane perpendicular to the magnetic field of the ALICE magnet and the e$^+$e$^-$ pair plane extrapolated 50~cm beyond the reconstructed conversion point.
An additional selection based on the cosine of the pointing angle with $\cos(\theta_\text{PA})>0.85$ is applied, where the pointing angle, $\theta_\text{ PA}$, is the angle between the reconstructed photon momentum vector and the vector joining the collision vertex and the conversion point.
A selection in the Armenteros-Podolanski plot~\cite{amenteros} which contains the distribution of $q_{\mbox{\tiny T}}=p_\text{daughter}\times\sin\theta_\mathrm{mother-daugther}$ versus the longitudinal momentum asymmetry ($\alpha=(p_\mathrm{L}^+-p_\mathrm{L}^-)/(p_\mathrm{L}^++p_\mathrm{L}^-)$) with $q_\text{T}<q_\text{T,max}\sqrt{1-\alpha^2/\alpha_\text{max}^2}$ where $q_\text{T,max}=0.05$~\GeVc\ and $\alpha_\text{max}=0.95$ removes the remaining contamination from K$^0_\text{S}$, $\Lambda$ and $\overline{\Lambda}$.
Additionally, as explained in \Ref{Abelev:2014ypa}, an out-of-bunch pileup correction is required for the PCM measurement, which estimates the contamination of photon candidates from multiple overlapping events in the TPC.
The correction is obtained from a study of the longitudinal distance of closest approach (DCA) of the conversion photon candidates which is the smallest distance in beam direction ($z$) between the primary vertex and the momentum vector of the photon candidate.
Photon candidates from different events generate a broad underlying Gaussian-like DCA distribution, which is described with a background estimator to describe the out-of-bunch pileup contribution. 
This correction is found to be transverse momentum dependent and ranges from 12\% at low \pT~($\approx0.5$ \GeVc) to 4\% at high \pT~($\approx7$ \GeVc) at both center-of-mass energies.

Photons and electrons/positrons produce electromagnetic showers as they enter an electromagnetic ca\-lo\-rimeter and their deposited energy can be measured.
By design, these showers usually spread over several adjacent calorimeter cells in the EMCal. 
Therefore, the reconstruction of the full energy of particles requires the grouping of such adjacent cells into clusters, for which a clusterization algorithm is used.
The cell with the highest deposited energy, exceeding a given seed energy, $E_{\rm seed}$, is used by the algorithm as a starting point.
The cluster is then formed by addition of all adjacent cells with individual energy above a minimum energy, $E_{\rm min}$.
This aggregation of cells continues as long as the energy of an adjacent cell is smaller than the energy of the previous cell.
Otherwise the clusterization algorithm stops the aggregation process.
The clustering procedure is repeated until all cells are grouped into clusters.
The energy deposited in the individual cells of the cluster is summed to obtain the total cluster energy.
For the presented \EMC\ analyses, the values of  $E_{\rm seed}=500$~MeV and $E_{\rm min}=100$~MeV are chosen, which are determined to suppress out-of-bunch background, as well as the general noise level of the front-end electronics.
Finally, a correction for the difference of relative energy scale and position of the EMCal between data and simulation is applied, which was obtained by reconstructing in data and simulation the average neutral pion mass peak as a function of the EMCal photon energy, pairing photon candidates from PCM with those of the EMCal~\cite{Acharya:2017hyu,Acharya:2017tlv}.

To select true photon candidates from the sample of reconstructed clusters, photon identification criteria are applied.
Clusters are required to have a minimum energy $E_{\rm cluster}>0.7$~GeV and should consist of at least two cells. 
EMCal clusters are accepted only if they are within $|\eta|<0.67$ and $1.40$~rad~$ < \varphi < 3.15$~rad.
A cluster timing selection relative to the collision time of $-35<t_{\rm cluster}<30$~ns at $\sqrt{s}=8$ TeV~($|t_{\rm cluster}|<50$ ns at $\sqrt{s}=2.76$ TeV) is imposed to remove pileup from multiple events that may occur within the readout interval of the front-end electronics. 
This constraint removes photon candidates from different bunch crossings with an efficiency of better than 99\%. 

Clusters, which may have a significant contribution from energy deposited by charged hadrons, are rejected by propagating charged particle tracks to the EMCal surface and associating them to clusters based on geometrical criteria, generally called ``track matching'' in what follows.
Track matching is applied in $\eta$ and $\varphi$ depending on track momentum, from $|\Delta\eta|<0.04$ and $|\Delta\varphi|<0.09$~rad for lowest \pT\ to $|\Delta\eta|<0.01$ and $|\Delta\varphi|<0.015$~rad at highest \pT.
Parameterized as $|\Delta\eta|<0.01+(\pT+4.07)^{-2.5}$ and $|\Delta\varphi|<0.015+(\pT+3.65)^{-2}$~rad, with \pT\ in units of GeV/$c$, these criteria result in a track matching efficiency of more than 95\% over the full \pT\ range.
Furthermore, the photon purity is significantly improved by the application of a cluster shape selection of $0.1<$ $\lzt < (0.32 + 0.0072\times E^2_{\text{clus}}/\text{GeV}^2)$ for $E_{\text{clus}}\leq5$ GeV and $0.1<$ $\lzt < 0.5$ for $E_{\text{clus}}>5$ GeV, used to suppress the contamination caused by overlapping clusters. 
Here, $\sigma^{2}_{\rm long}$ stands for the larger eigenvalue of the dispersion matrix of the shower shape ellipse defined by the corresponding cell indices in the supermodule and their energy contributions to the cluster~\cite{Awes1992130,Acharya:2017hyu}.
In addition, by applying $\sigma^{2}_{\rm long}>0.1$ the contamination caused by neutrons hitting the APDs of the readout electronics is removed.

Corrections for reconstruction efficiencies, conversion probability and purity are evaluated using the PYTHIA8~\cite{Sjostrand:2014zea} and PHOJET~\cite{Engel:1995sb} MC event generators.
Particles generated by the event generator are propagated through the ALICE detector using GEANT3~\cite{geant3ref2}. 
The same reconstruction algorithms and analysis selection ranges are applied as those in data.
The correction factors for both MC productions are found to be consistent, and are therefore combined to reduce the statistical uncertainties.
Before the efficiency correction, $\pT$-scale and resolution effects are corrected using Bayesian unfolding~\cite{DAGOSTINI1995487} with the detector response is used to convert from the reconstructed to the true \pT\ of the photons. 
Consequently, the reconstruction efficiency is calculated as a function of the true transverse momentum by dividing the reconstructed Monte Carlo~(MC) validated photon spectrum by all photons from the simulation.
The reconstruction efficiency is found to be largest at $\pT\approx3$~\GeVc\ with 73\% for PCM and 56\% at $\pT\approx5$ \GeVc\ for \EMC\ in the respective detector acceptance, decreasing with lower and higher \pT\ for both methods.
For the photons reconstructed with PCM, a further correction based on MC information is applied to account for the conversion probability of the photons in the detector material, which increases from 5.6\% at the lowest to 8.9\% at the highest measured \pT, mainly due to the minimum electron track momentum requirement.

A correction based on MC information for the contamination of the photon sample from falsely identified and subsequently combined tracks of electrons, pions, kaons or muons is applied for the conversion method.
The purity of the photon sample reconstructed with PCM is found to be 99\% up to 3 \GeVc\ and decreases down to 96\% at high transverse momentum due to the increasing contamination from electron-pion pairs.
For the calorimeter-based method, a similar purity correction is applied but for falsely identified photon candidates mainly from clusters created by neutrons and antineutrons at low \pT\ and by neutral kaons at high \pT.
The purity correction is the largest for $\pT<3$ \GeVc\ where purities between 87\% and 97\% rising with \pT\ were found, while at high \pT\ the purities reach values of 97\%.

For the inclusive photon measurements, contributions of secondary photons from weak decays and hadronic interactions are estimated and removed.
The main source of photons from weak decays are K$^0_\text{S}$ decays, however contributions from K$^0_\text{L}$ and $\Lambda$ are also considered.
The correction uses the decay photon cocktail simulation described in \Sect{sec:cocktail} which provides the secondary photon yields.
Taking into account the detector response and detection efficiency for the different reconstruction techniques, these photons are removed from the photon sample.
The remaining correction factor for secondary photons, for example due to interactions with the detector material, are obtained purely from MC information.
The secondary corrections are of the order of 1--3\% for K$^0_\text{S}$, $\lesssim$ 0.05--0.2\% for K$^0_\text{L}$, $\lesssim0.02\%$ for $\Lambda$ and 0.1--2.5\% for material interactions depending on \pT\ and on the photon reconstruction technique within the given ranges.
In general the correction factors tend to be larger for the \EMC\ reconstruction technique, due to the worse pointing resolution of the photons.

The neutral pion and $\eta$ meson measurements, which are needed to extract $\Rg$ from \Eq{eq:Rg}, are described in detail in \Refs{Acharya:2017hyu,Acharya:2017tlv}.
The meson yields are obtained for PCM, \EMC\ and a hybrid method~(\PCMEMC), in which photon candidates reconstructed with PCM are paired with those reconstructed in the EMCal.
For the measurement of $\Rg$  with the \PCMEMC\ method, it is beneficial to measure the inclusive photons with PCM.
However, to be consistent with the corresponding meson measurements~\cite{Acharya:2017tlv}, a wider selection range of $-4<\text{n}_{\sigma_e}<5$ on the energy loss hypothesis of the electron/positron in the $\sqrt{s}=8$ TeV measurement is used, and the charged pion d$E$/d$x$ based rejection is applied independent of \pT\ in both collision systems for the corresponding inclusive photon measurement with PCM.

\section{Systematic uncertainties}
\label{sec:systematics}

Systematic uncertainties are summarized for the measurements of $\Ygi$ and $\Rg$ in \Tab{tab:table_system_2760} for $\sqrt{s}=2.76$~TeV and in \Tab{tab:table_system_8} for $\sqrt{s}=8$ TeV and shown for three transverse momentum bins used in the analyses.
The uncertainties are given in percent and for each reconstruction method individually.
The detailed description of uncertainties related to the $\pi^0$ meson measurements that enter into the calculation of the direct photon excess ratios $\Rg$ can be found in~\cite{Acharya:2017hyu} for $\sqrt{s}=2.76$ TeV and in~\cite{Acharya:2017tlv} for $\sqrt{s}=8$ TeV.
All uncertainties are evaluated on the fully corrected spectra of $\Ygi$ or directly on $\Rg$.
In case of $\Rg$, the systematic uncertainties therefore also contain the effects of the systematic variations on the measured neutral pion spectrum, thus benefiting from partial cancellations of common uncertainties.

\begin{table}[t!]
  \begin{center}
  \vspace{-0.3cm}
    \begin{tabular}{l||cc||cc|c|cc||cc|c|cc}
      \hline
      \pT\ interval (\GeVc)& \multicolumn{2}{c||}{$0.4-0.6$} & \multicolumn{5}{c||}{$1.6-1.8$ } & \multicolumn{5}{c}{$6.0-8.0$ }  \\ \hline 
      \multirow{1}{*}{Method} & \multicolumn{2}{c||}{\multirow{1}{*}{PCM}} & \multicolumn{2}{c|}{\multirow{1}{*}{PCM}} & \multicolumn{1}{c|}{P-E} & \multicolumn{2}{c||}{\multirow{1}{*}{EMC}}  & \multicolumn{2}{c|}{\multirow{1}{*}{PCM}} & \multicolumn{1}{c|}{P-E} & \multicolumn{2}{c}{\multirow{1}{*}{EMC}}\\ 
      Measurement& $\Ygi$ & $\Rg$& $\Ygi$ & $\Rg$& $\Rg$& $\Ygi$ & $\Rg$& $\Ygi$ & $\Rg$& $\Rg$& $\Ygi$ & $\Rg$ \\ \hline \hline
      Inner material         & $4.5$ & $4.5$ & $4.5$ & $4.5$ &  --   &  --   &  --   & $4.5$ & $4.5$ &  --   &  --   &   --    \\
      Outer material         &  --   &  --   &  --   &  --   & $2.1$ & $2.1$ & $3.0$ &  --   &  --   & $2.1$ & $2.1$ & $3.0$  \\
      PCM track rec.         & $0.3$ & $3.3$ & $0.3$ & $1.6$ & $1.3$ &  --   &  --   & $0.3$ & $8.4$ & $1.3$ &  --   &   --    \\
      PCM electron PID       & $0.5$ & $1.4$ & $0.6$ & $1.8$ & $0.4$ &  --   &  --   & $1.3$ & $13.4$& $3.8$ &  --   &   --    \\
      PCM photon PID         & $0.4$ & $5.4$ & $0.6$ & $2.5$ & $1.1$ &  --   &  --   & $2.2$ & $11.4$& $3.1$ &  --   &   --    \\
      Cluster description    &  --   &  --   &  --   &  --   & $2.6$ & $2.7$ & $4.1$ &  --   &  --   & $5.5$ & $2.7$ & $4.0$  \\
      Cluster energy calib.  &  --   &  --   &  --   &  --   & $2.0$ & $1.4$ & $2.0$ &  --   &  --   & $2.6$ & $2.0$ & $2.5$  \\
      Track match to cluster &  --   &  --   &  --   &  --   & $1.5$ & $0.7$ & $0.7$ &  --   &  --   & $5.7$ & $0.7$ & $1.4$  \\
      Efficiency             &  --   &  --   &  --   &  --   & $2.0$ & $1.5$ & $2.5$ &  --   &  --   & $2.0$ & $1.5$ & $2.5$  \\
   Signal extraction $\pi^0$ &  --   & $5.0$ &  --   & $2.7$ & $2.1$ &  --   & $2.5$ &  --   & $4.9$ & $3.7$ &  --   & $2.4$  \\
      Cocktail               &  --   & $0.9$ &  --   & $2.2$ & $1.3$ &  --   & $1.4$ &  --   & $3.4$ & $3.1$ &  --   & $2.3$  \\
      Pileup                 & $2.4$ & $2.6$ & $1.1$ & $1.2$ & $0.9$ & $0.3$ &  --   & $1.6$ & $1.7$ & $0.9$ & $0.3$ &   --    \\
      \hline
      Total syst. uncertainty& $5.1$ & $9.7$ & $4.7$ & $6.7$ & $5.6$ & $4.0$ & $6.7$ & $5.4$ & $20.9$& $11.3$& $4.3$ & $7.1$  \\
      Statistical uncertainty& $0.2$ & $7.8$ & $0.7$ & $4.3$ & $4.4$ & $0.5$ & $5.0$ & $6.3$ & $23.1$& $18.3$& $4.6$ & $8.6$  \\
      \hline \hline
      Measurement& $\Ygi$ & $\Rg$& \multicolumn{2}{c}{$\Ygi$} & \multicolumn{3}{c||}{$\Rg$} & \multicolumn{2}{c}{$\Ygi$} & \multicolumn{3}{c}{$\Rg$}\\ \hline \hline
      Comb syst. uncertainty& $5.1$  & $9.7$ & \multicolumn{2}{c}{$3.1$} & \multicolumn{3}{c||}{$4.6$} & \multicolumn{2}{c}{$2.9$} & \multicolumn{3}{c}{$7.1$}  \\ 
      Comb stat. uncertainty& $0.2$  & $7.8$ & \multicolumn{2}{c}{$0.4$} & \multicolumn{3}{c||}{$2.7$} & \multicolumn{2}{c}{$3.9$} & \multicolumn{3}{c}{$7.2$}  \\ \hline
    \end{tabular}
    \caption{Summary of relative systematic uncertainties in percent for selected \pT\ bins for the reconstruction of inclusive photons and the $\Rg$ measurement at $\sqrt{s}=2.76$ TeV.
    The hybrid method \PCMEMC\ is abbreviated as \mbox{P-E} in this table.
            The statistical uncertainties are given in addition to the total systematic uncertainty as well as the uncertainties after combination of the independent measurements.
            The visible cross section uncertainty for $\sigma_{\rm MB_{OR}}$ of $2.5\%$ is independent from reported measurements and is separately indicated in the figures below. }
    \label{tab:table_system_2760}
  \end{center}
\end{table}

\begin{table}[t!]
  \begin{center}
  \vspace{0.1cm}
    \begin{tabular}{l||cc||cc|c|cc||cc|c|cc}
      \hline
      \pT\ interval (\GeVc)& \multicolumn{2}{c||}{$0.4-0.6$} & \multicolumn{5}{c||}{$1.6-1.8$ } & \multicolumn{4}{c}{$9.0-12.0$ }  \\ \hline 
      \multirow{1}{*}{Method} & \multicolumn{2}{c||}{\multirow{1}{*}{PCM}} & \multicolumn{2}{c|}{\multirow{1}{*}{PCM}} & \multicolumn{1}{c|}{P-E} & \multicolumn{2}{c||}{\multirow{1}{*}{EMC}}  & \multicolumn{2}{c|}{\multirow{1}{*}{PCM}} & \multicolumn{1}{c|}{P-E} & \multicolumn{2}{c}{\multirow{1}{*}{EMC}}\\  
      Measurement& $\Ygi$ & $\Rg$& $\Ygi$ & $\Rg$& $\Rg$& $\Ygi$ & $\Rg$& $\Ygi$ & $\Rg$& $\Rg$& $\Ygi$ & $\Rg$ \\ \hline\hline
      Inner material         & $4.5$ & $4.5$ & $4.5$ & $4.5$ &  --   &  --   &  --   & $4.5$ & $4.5$ &  --   &  --   &   --    \\
      Outer material         &  --   &  --   &  --   &  --   & $2.1$ & $2.1$ & $3.0$ &  --   &  --   & $2.1$ & $2.1$ & $3.0$  \\
      PCM track rec.         & $0.2$ & $0.5$ & $0.1$ & $0.5$ & $0.2$ &  --   &  --   & $0.1$ & $0.5$ & $0.2$ &  --   &   --    \\
      PCM electron PID       & $1.1$ & $2.4$ & $0.6$ & $0.8$ & $0.3$ &  --   &  --   & $0.7$ & $0.8$ & $0.8$ &  --   &   --    \\
      PCM photon PID         & $1.8$ & $1.2$ & $1.3$ & $1.0$ & $1.0$ &  --   &  --   & $2.3$ & $5.5$ & $1.7$ &  --   &   --    \\
      Cluster description    &  --   &  --   &  --   &  --   & $2.5$ & $2.6$ & $3.0$ &  --   &  --   & $3.0$ & $2.6$ & $1.9$  \\
      Cluster energy calib.  &  --   &  --   &  --   &  --   & $2.3$ & $1.4$ & $2.3$ &  --   &  --   & $1.8$ & $0.9$ & $1.8$  \\
      Track match to cluster &  --   &  --   &  --   &  --   & $0.2$ & $1.8$ & $1.5$ &  --   &  --   & $1.9$ & $1.8$ & $1.6$  \\
      Efficiency             & $0.5$ & $0.5$ & $0.5$ & $0.5$ & $2.1$ & $1.8$ & $2.7$ & $0.5$ & $0.5$ & $2.1$ & $1.8$ & $2.7$  \\
   Signal extraction $\pi^0$ &  --   & $4.9$ &  --   & $1.6$ & $1.8$ &  --   & $2.7$ &  --   & $6.6$ & $3.1$ &  --   & $1.9$  \\
      Cocktail               & $0.2$ & $1.7$ & $0.1$ & $0.7$ & $1.0$ & $0.3$ & $0.8$ & $0.1$ & $0.5$ & $1.9$ & $0.3$ & $1.4$  \\
      Pileup                 & $3.8$ & $4.3$ & $2.7$ & $4.2$ & $2.7$ & $0.1$ &  --   & $4.3$ & $4.4$ & $3.0$ & $0.1$ &   --   \\
      \hline
      Total syst. uncertainty& $6.3$ & $8.5$ & $5.4$ & $6.6$ & $5.7$ & $4.4$ & $6.4$ & $6.4$ & $10.7$& $7.1$ & $4.3$ & $5.6$  \\
      \hline\hline
      Statistical uncertainty& $0.1$ & $4.3$ & $0.3$ & $2.2$ & $2.1$ & $0.2$ & $2.7$ & $3.3$ & $17.2$& $9.9$ & $2.1$ & $4.7$  \\
      \hline \hline
      Measurement& $\Ygi$ & $\Rg$& \multicolumn{2}{c}{$\Ygi$} & \multicolumn{3}{c||}{$\Rg$} & \multicolumn{2}{c}{$\Ygi$} & \multicolumn{3}{c}{$\Rg$}\\ \hline \hline
      Comb syst. uncertainty& $6.3$ & $8.5$ & \multicolumn{2}{c}{$3.5$} & \multicolumn{3}{c||}{$4.5$} & \multicolumn{2}{c}{$3.6$} & \multicolumn{3}{c}{$5.9$}  \\ 
      Comb stat. uncertainty& $0.1$ & $4.3$ & \multicolumn{2}{c}{$0.2$} & \multicolumn{3}{c||}{$1.4$} & \multicolumn{2}{c}{$1.8$} & \multicolumn{3}{c}{$4.3$}  \\ \hline
    \end{tabular}
    \caption{Summary of relative systematic uncertainties in percent for selected \pT\ bins for the reconstruction of inclusive photons and the $\Rg$ measurement at $\sqrt{s}=8$ TeV.
    The hybrid method \PCMEMC\ is abbreviated as \mbox{P-E} in this table.
            The statistical uncertainties are given in addition to the total systematic uncertainty as well as the uncertainties after combination of the independent measurements.
            The visible cross section uncertainty of $2.6\%$ is independent from reported uncertainties and is separately indicated in the figures below. }
    \label{tab:table_system_8}
  \end{center}
\end{table}

For the PCM measurements, the material budget uncertainty is the main contributor to the total uncertainty and its value of 4.5\% was previously determined in~\cite{Abelev:2012cn,Abelev:2014ffa}.
Systematic uncertainties associated with ``track reconstruction'' are the uncertainties that are estimated from variations of required TPC clusters as well as minimum transverse momentum requirements of tracks.
Particle identification (PID) uncertainties are determined by variation of the PID selection ranges of electrons and photons as described in \Sect{sec:photonreco}.
A systematic uncertainty is estimated for the pileup corrections that are applied in the analyses.
It is dominated by the contribution from the DCA background description for the out-of-bunch pileup estimation but also contains the uncertainty from the SPD in-bunch pileup rejection due to its limited efficiency.

The systematic uncertainty of the \EMC\ measurement contains a large contribution from the limited knowledge of the outer material budget, which is composed by all detector components from the radial center of the TPC up to the EMCal. 
This uncertainty is determined by comparing the effects on the corrected spectra using inputs from data taking campaigns with and without TRD modules in front of the EMCal.
This could be done as the EMCal was masked only partially by the TRD during the data taking in 2011 and 2012.
The material budgets of TRD and TOF are roughly similar and therefore the quoted uncertainty is taken as $\sqrt{2}$ times the difference of the corrected spectra with and without TRD modules in front of the EMCal.
Systematic uncertainties contributing to the ``cluster description'' category are the uncertainties associated to the description of clusters in simulation which influence the reconstruction efficiencies.
The associated variables are the minimum cluster energy, shower shape, number of cells, time and clusterization seed as well as minimum energy selection variations.
The uncertainty of non-linearity effects as well as the energy scale of clusters are incorporated in the ``cluster energy calibration''.
To assess this uncertainty different parametrizations for the MC $\pi^0$ mass peak position correction are considered to account for the residual differences between data and MC.
The ``efficiency'' uncertainty reflects the differences between the MC generators that are used for the efficiency calculation.
The pileup systematic uncertainty reflects the finite efficiency of the SPD for in-bunch pileup rejection.

The hybrid method \PCMEMC\ requires the same evaluation of uncertainties as its individual standalone methods.
However, most systematics show a different size or behavior on $\Rg$ as the contained inclusive photon measurement is PCM-based whereas for the neutral pion one photon candidate of each reconstruction approach is used.
In addition, the ``track matching to cluster'' uncertainty includes the uncertainties associated with the matching of $\Vz$ tracks or primary tracks with the cluster, which is an important ingredient for the hybrid method.

The uncertainty on the decay photon simulation is obtained by varying the parametrizations of the neutral pion and $\eta$ meson for each reconstruction technique within the \pT-uncorrelated systematic and statistical uncertainties. 
This leads to an associated uncertainty of $0.9-3$\% and $0.5-2\%$ for \pp\ collisions at $\sqrt{s}=2.76$ and $8$~TeV, respectively, which strongly depends on \pT.
Furthermore, a variation of the \mT\ scaling constants has been considered for the remaining mesons which yields an uncertainty below 0.1\%.

Partial systematic uncertainty cancellations are present for the direct photon excess ratio $\Rg$.
The material budget uncertainty in the PCM measurement, which enters once in the inclusive photon measurement and twice in the neutral pion measurement, cancels once in $\Rg$.
A similar cancellation is present in the \EMC\ measurement, where the outer material budget uncertainty cancels partially in the double ratio as well.
For the hybrid method, the inner material budget uncertainty cancels fully in the double ratio and only the outer material budget uncertainty enters once in the total uncertainty, which is the main advantage of using this reconstruction method for $\Rg$.
 
The final estimated systematic uncertainties on the inclusive photon cross section amount to 5--7\% for the conversion method and 4--9\% for the \EMC\ measurement in the measured $\pT$ range. 
The material budget uncertainty of the conversion method is the dominant source, whereas the calorimeter-based method shows a strong dependence on the cluster description in the simulation and the associated efficiency estimates.
With statistical uncertainties below 1\% for $\pT<3$ \GeVc\ the inclusive photon measurement is therefore limited by the systematic uncertainties.

The systematic uncertainties on the direct photon excess ratio $\Rg$ are larger than for the inclusive photons due to the addition of the neutral pion related uncertainties.
For PCM, the systematic uncertainties amount to 6--20\% dominated by the material budget uncertainty and the neutral pion signal extraction uncertainties at low and high \pT.
Systematic uncertainties for the \EMC\ measurements are smaller at high transverse momentum compared to PCM with values of 7--9\% at $\sqrt{s}=2.76$ TeV and 6--8\% at $\sqrt{s}=8$ TeV with dominant contributions of the outer material budget, the cluster description and neutral pion signal extraction uncertainties.
Mostly due to the cancellation of the inner material budget uncertainty, the hybrid method \PCMEMC\ exhibits the smallest systematic uncertainty at intermediate \pT\ with values of 6\% at $\sqrt{s}=2.76$ TeV and 5.4\% at $\sqrt{s}=8$ TeV.

\section{Results}
\label{sec:results}
\begin{figure}[t]
        \center
        \includegraphics[width=.49\textwidth]{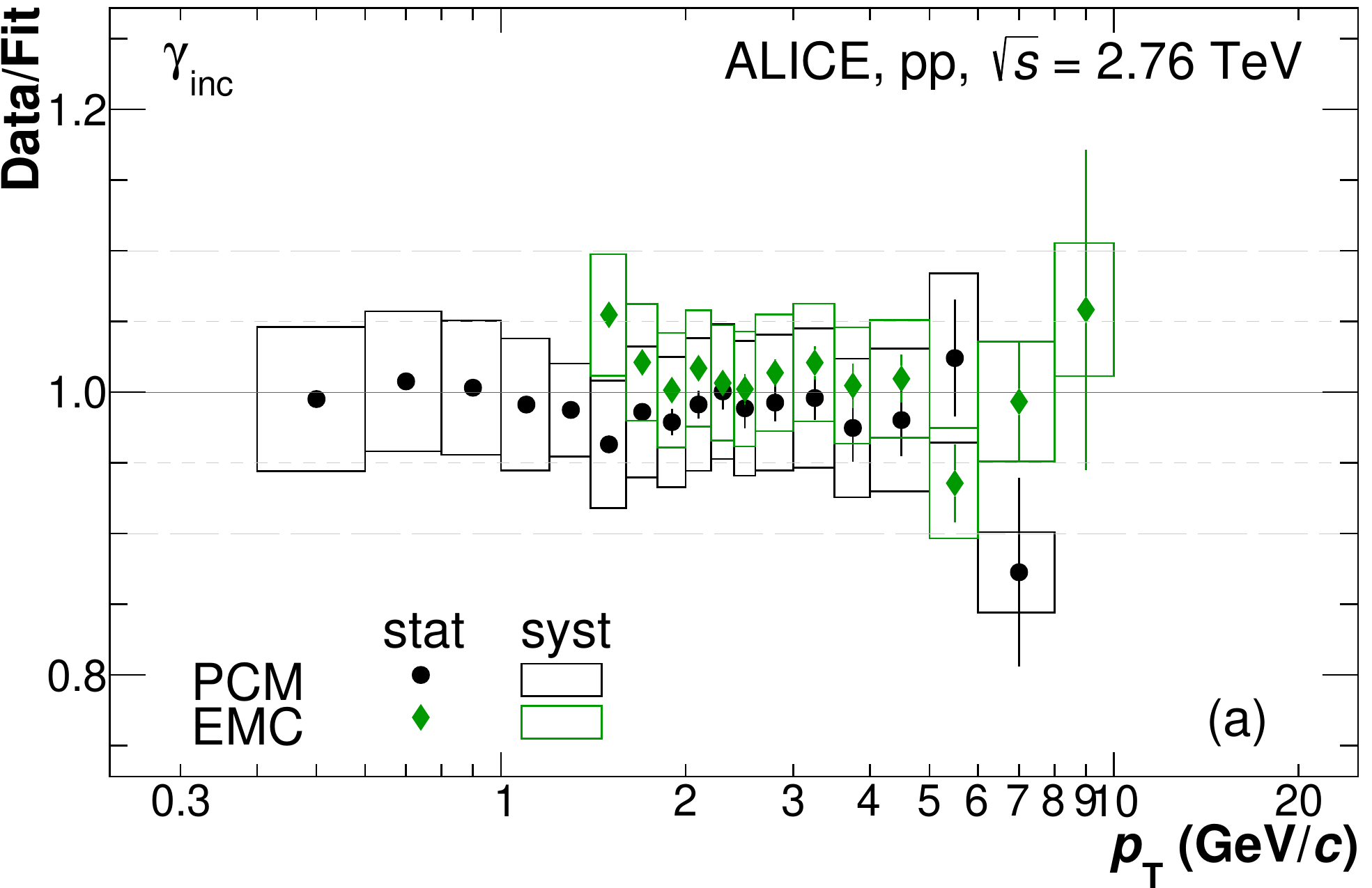}
        \hspace{0.1cm}
        \includegraphics[width=.49\textwidth]{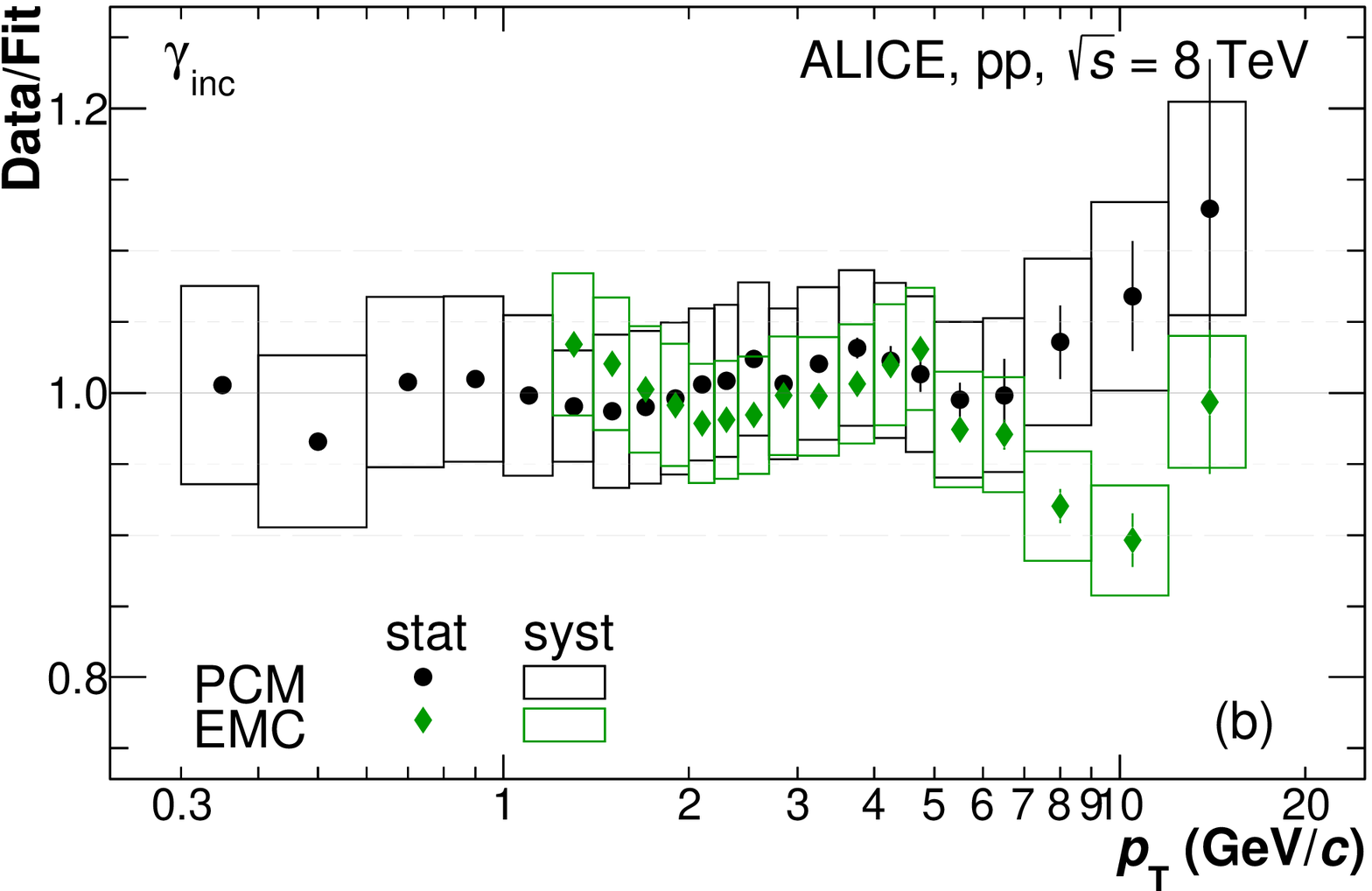}
        \caption{Ratio of inclusive photon invariant cross sections obtained with the PCM and \EMC\ methods to the TCM fit, \Eq{eq:Bylinkin} of the combined inclusive photon spectrum in pp collisions at $\sqrt{s}=2.76$ TeV~\textit{(a)} and 8 TeV~\textit{(b)}.
                 Statistical uncertainties are given by vertical lines and systematic uncertainties are visualized by the height of the boxes, while the bin widths are represented by the widths of the boxes.}
        \label{fig:incgammaratios}
\end{figure}

The invariant cross sections for inclusive photons at mid-rapidity~($|y|<0.9$) are given as
\begin{equation}
 E\frac{\text{d}^3\sigma^{pp\rightarrow\gamma+X}}{\text{d}p^3}=\frac{1}{2\pi\pT}\frac{1}{\mathscr{L_{\rm int}}}\frac{\epsilon_\text{pur}}{P_\text{conv}\epsilon_\text{rec}A}\frac{F_\text{pile-up}\cdot N^\gamma-N^\gamma_\text{sec}}{\Delta y\Delta\pT},
\end{equation}
where $\epsilon_\text{pur}$, $P_\text{conv}$ and $\epsilon_\text{rec}$ are the purity, conversion probability and reconstruction efficiency correction factors, respectively, and $\mathscr{L_{\rm int}}$ is the integrated luminosity.
The conversion probability as well as the out-of-bunch pileup correction factor ($F_\text{pile-up}$) only apply for the PCM measurement.
The acceptance correction factor, $A$, is only applied for EMCal to account for the limited azimuth coverage.
In addition, the inclusive photon raw yield is given by $N^\gamma$ and the summed secondary photon raw yields by $N^\gamma_\text{sec}$.
Furthermore, the interval ranges in rapidity and transverse momentum are given by $\Delta y \Delta\pT$.

The double ratios are measured by combining the individual inclusive photon and neutral pion spectra from the same reconstruction methods with a cocktail simulation based on the same neutral pion spectrum.
In this way, possible biases can be removed, since they would affect both the inclusive photon and the neutral pion measurements.

\begin{figure}[t]
        \center
        \includegraphics[width=.49\textwidth]{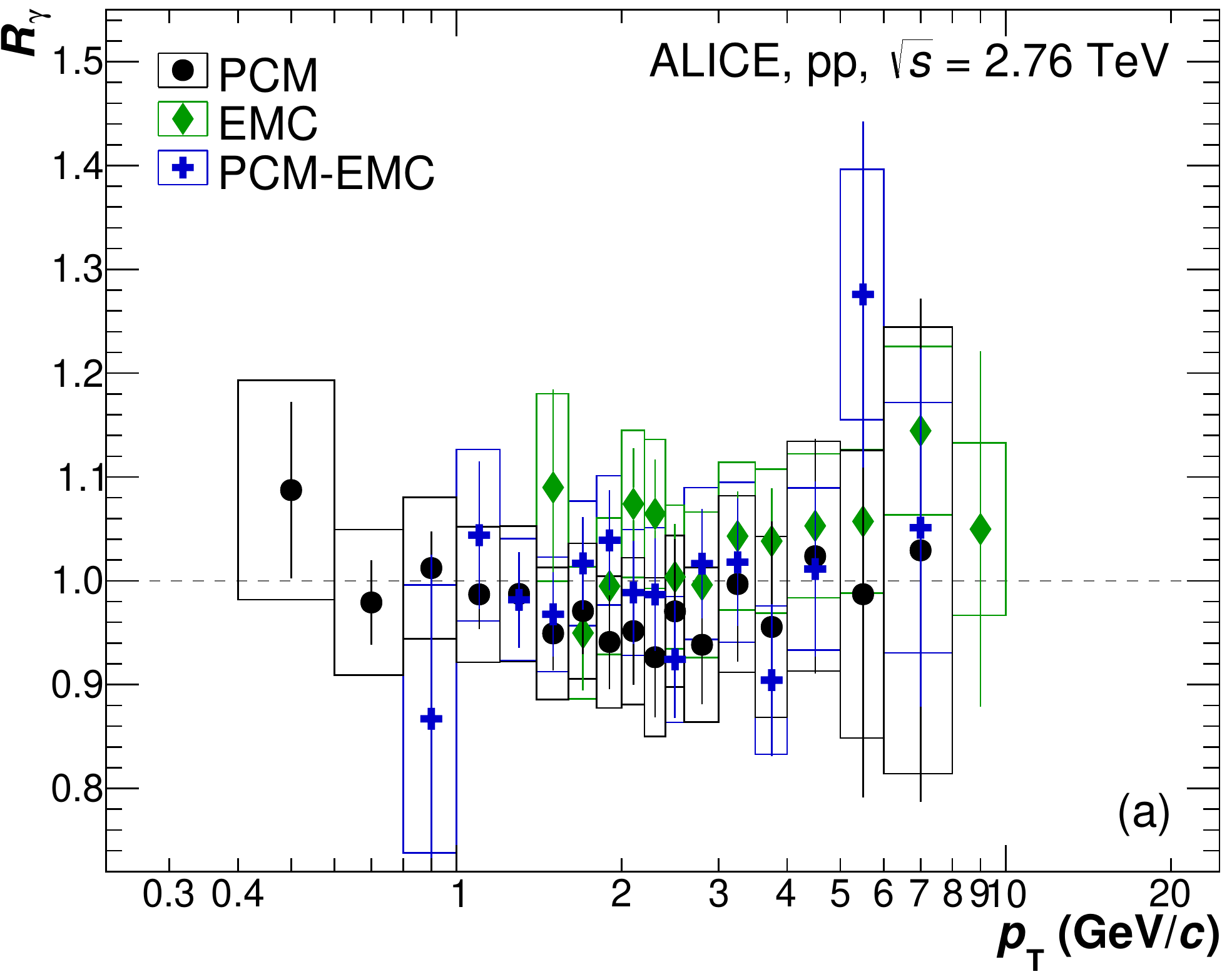}
        \hspace{0.1cm}
        \includegraphics[width=.49\textwidth]{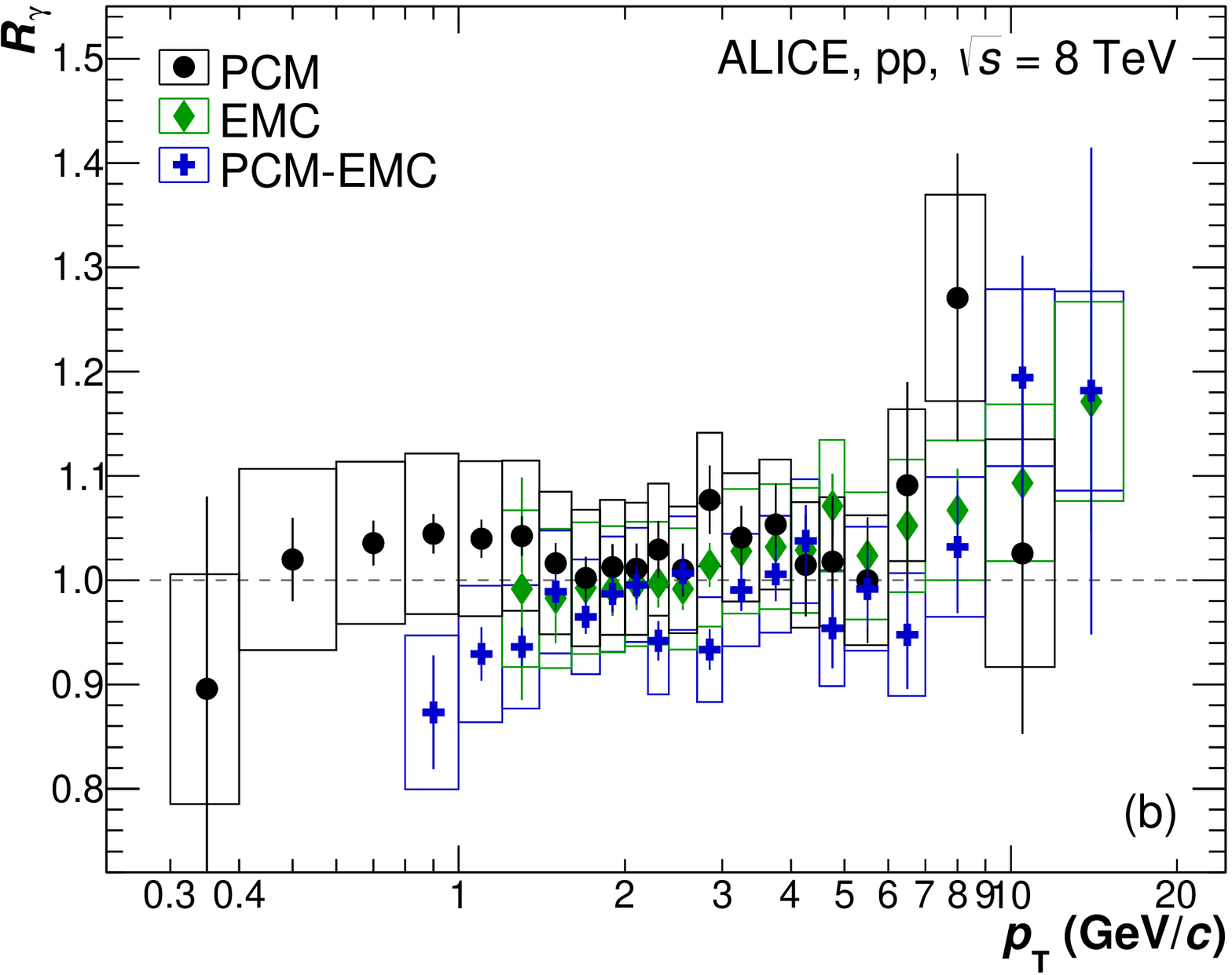}
        \caption{Direct photon excess ratios obtained with the PCM, \EMC\ and \PCMEMC\ methods in pp collisions at $\sqrt{s}=2.76$ TeV~\textit{(a)} and 8 TeV~\textit{(b)}.
                 Statistical uncertainties are given by vertical lines and systematic uncertainties are visualized by the height of the boxes, while the bin widths are represented by the widths of the boxes.}
        \label{fig:DRindmeas}
\end{figure}
\begin{figure}
        \center
        \includegraphics[width=.49\textwidth]{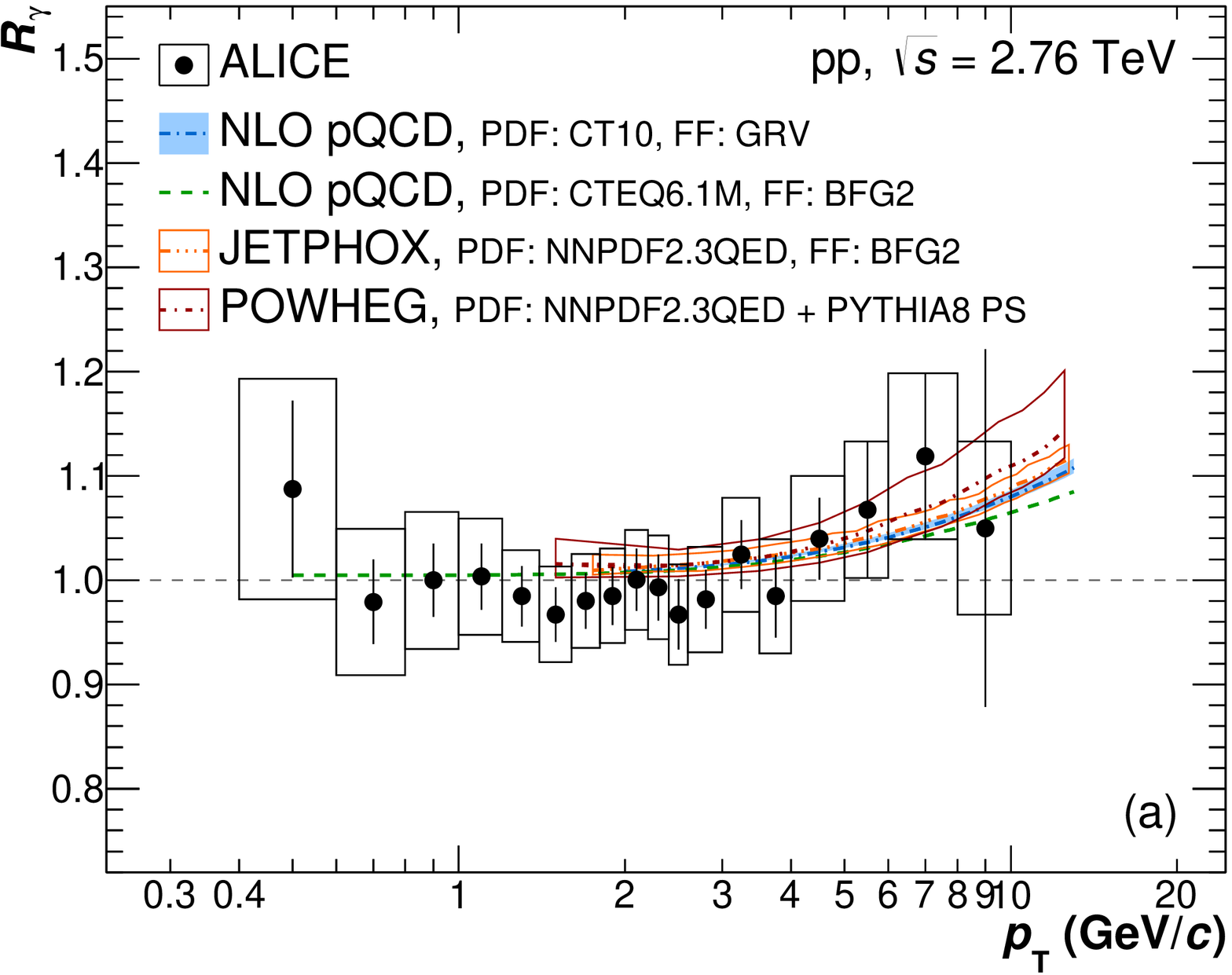}
        \hspace{0.1cm}
        \includegraphics[width=.49\textwidth]{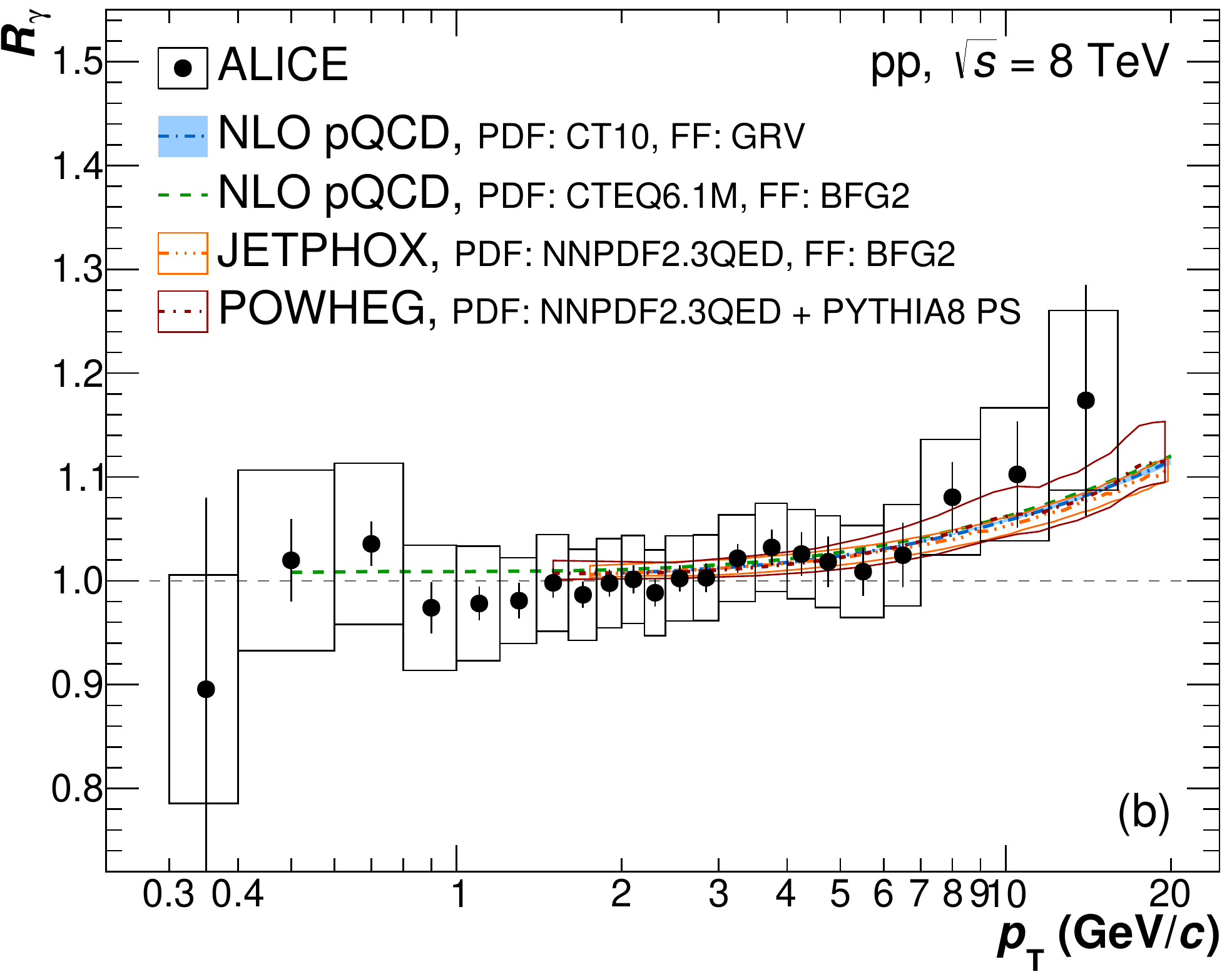}
        \caption{Direct photon excess ratios for the combined measurements at $\sqrt{s}=2.76$~\textit{(a)} and 8 TeV~\textit{(b)} including pQCD NLO predictions with CT10~\cite{Lai:2010vv,Gao:2013xoa,Guzzi:2011sv} or CTEQ6.1M~\cite{Stump:2003yu} proton PDF and GRV~\cite{Gluck:1992zx} or BFG2~\cite{Bourhis:1997yu} FF.
        In addition, a JETPHOX calculation~\cite{Klasen:2017dsy} based on NNPDF2.3QED~\cite{Ball:2013hta} proton PDF and BFG2 FF as well as a POWHEG calculation~\cite{Klasen:2017dsy} based on the same PDF but with the PYTHIA 8 parton shower algorithm are provided.}
        \label{fig:DRcombtheory}
\end{figure}

The individually measured inclusive photon invariant differential cross sections of the PCM and \EMC\, as well as the double ratios of the PCM, \PCMEMC\ and \EMC\ reconstruction methods are combined to obtain the final spectra and double ratios, respectively.
For the combination, the ``Best Linear Unbiased Estimates'' (BLUE) method~\cite{barlow1989statistics,Lyons1988rp,Valassi2003mu,Lyons1986em,Valassi2013bga} with full treatment of statistical and systematic uncertainty correlations was used.
For the inclusive photon measurement, the \EMC\ measurement is assumed to be fully independent of the PCM measurement both statistically and systematically.
However, for $\Rg$ the statistical uncertainties show partial correlation between the \PCM\ and \PCMEMC\ which are determined to be $\sim 20-50\%$ depending on \pT, since both measurements are based on the \PCM\ inclusive photon measurement using different subsets of the data; however, the statistical uncertainties of the neutral pion measurements are fully independent due to their different reconstruction methods.
The systematic uncertainty correlations were approximated via \pT\ dependent correlation factors.
It has been found that the largest correlations of the systematic uncertainties are among the \PCMEMC\ and the \PCM\ or \EMC\ methods, respectively. 
The fraction of correlation among the systematic uncertainties of the \PCMEMC\ and \EMC\ method has been estimated to be between 60--80\%, which can be attributed to the common uncertainty regarding the cluster reconstruction and efficiency uncertainties as well as the outer material budget.
For the \PCM\ and \PCMEMC\ methods the uncertainties regarding the \PCM\ photon identification and selection are largely correlated and thus the correlation factor ranges between 45--70\% depending on transverse momentum.

\begin{figure}[t]
        \center
        \includegraphics[width=.49\textwidth]{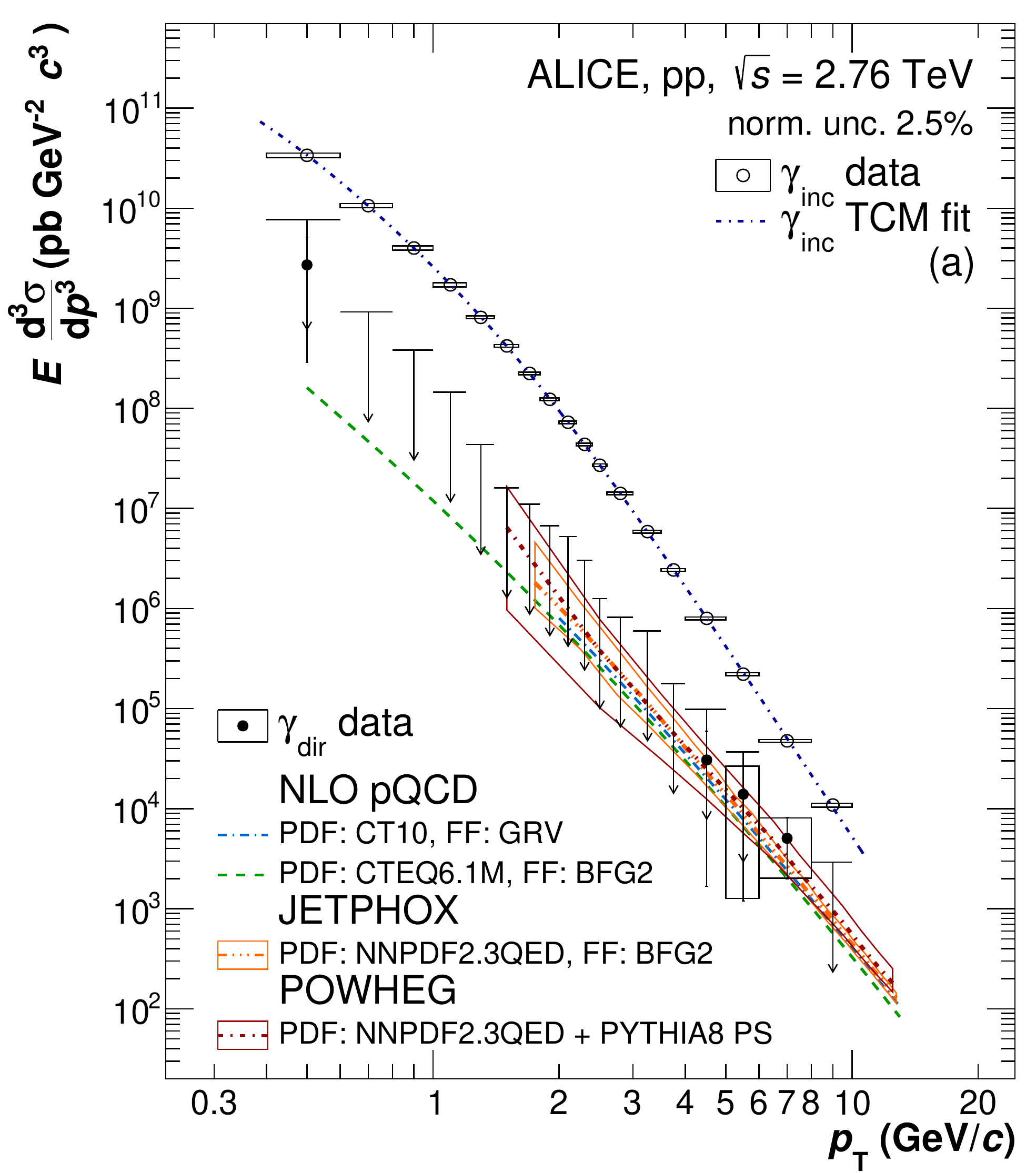}
        \includegraphics[width=.49\textwidth]{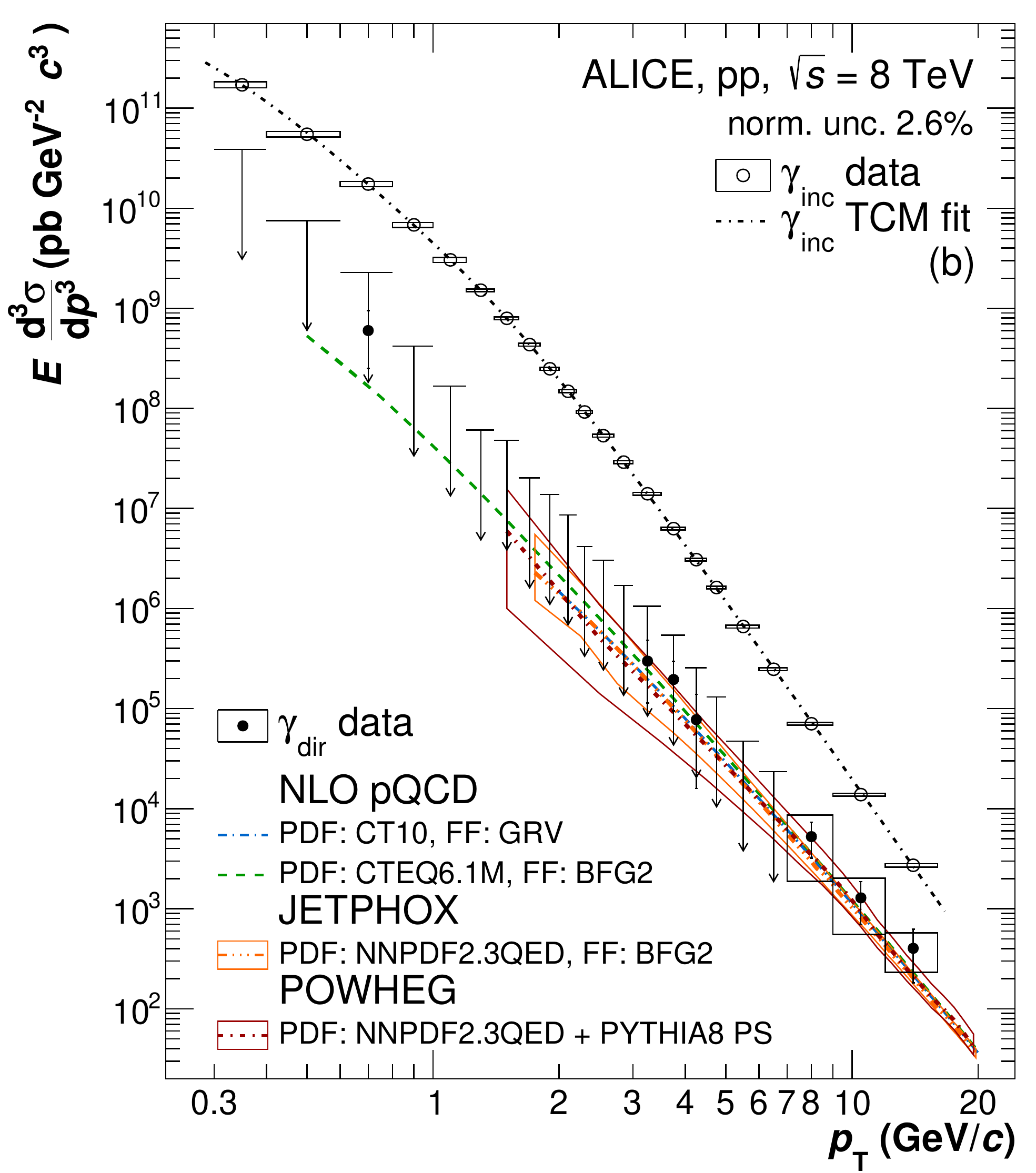}
        \caption{Upper limits of direct photon production at 90\% C.L. together with the invariant inclusive photon cross section at 2.76~\textit{(a)} and 8 TeV~\textit{(b)} including pQCD NLO predictions with CT10~\cite{Lai:2010vv,Gao:2013xoa,Guzzi:2011sv} or CTEQ6.1M~\cite{Stump:2003yu} proton PDF and GRV~\cite{Gluck:1992zx} or BFG2~\cite{Bourhis:1997yu} FF.
        In addition, a JETPHOX calculation~\cite{Klasen:2017dsy} based on NNPDF2.3QED~\cite{Ball:2013hta} proton PDF and BFG2 FF as well as a POWHEG~\cite{Klasen:2017dsy} calculation based on the same PDF but with the PYTHIA 8 parton shower algorithm are provided.
                 Upper limits are calculated for all spectrum points where the $\Rg$ value including its total uncertainty is consistent with unity and indicated by the horizontal bars at the end of the arrows in the figure.
                 Furthermore, direct photon cross sections are given where $\Rg$ is larger than unity considering only statistical uncertainties (round markers and vertical error bars) or taking only systematic uncertainties into account (boxes). 
                 }
        \label{fig:incgammadirgamma}
\end{figure}

The combined invariant cross sections of inclusive and direct photons, as well as the direct photon excess ratios $\Rg$, cover transverse momentum ranges of $0.4<\pT<10$ \GeVc\ and $0.3<\pT<16$ \GeVc\ for $\sqrt{s}=2.76$ and 8 TeV, respectively.
The combined inclusive photon spectra are shown in \Fig{fig:incgammadirgamma} together with a two-component model (TCM) fit~\cite{Bylinkin:2015xya}, whose functional form is a combination of an exponential function at low \pT\ and a power-law at high \pT, given as
\begin{eqnarray}
  E\frac{{\rm d}^{3}\sigma}{{\rm d}p^{3}} = A_{\rm e}\,\exp\left(\frac{-\pT}{T_{\rm e}}\right)+A\left( 1+\frac{\pT^{2}}{T^{2}n}\right)^{-n},
  \label{eq:Bylinkin}
\end{eqnarray}
with the free parameters $A_{\rm e}$, $A$, $T_{e}$, $T$ and $n$.
The two-component model is fitted to the inclusive photon spectra by using the total uncertainties of the spectra, obtained by quadratic combination of statistical and systematic uncertainties.
It is used only to facilitate a comparison of the methods in the ratio to the fit. 

The ratios of the inclusive photon spectra measured individually by PCM and \EMC\ relative to the TCM fit are shown in \Fig{fig:incgammaratios}, demonstrating that the inclusive spectra measured with PCM and \EMC\ agree within the uncertainties. 
By combining the two independent reconstruction techniques the systematic uncertainties are decreased to about $3$--$3.5\%$ between $1.4$ and $8$ \GeVc, while the statistical uncertainties are mostly below $1\%$.
The individual double ratios of the three reconstruction methods in both systems are displayed in \Fig{fig:DRindmeas} and the combined double ratios are shown in \Fig{fig:DRcombtheory}.
Through the combination of the three partially independent methods the systematic uncertainty could be reduced to about $4.5$--$5.5\%$ between $1$ and $3$ \GeVc\ for both $\sqrt{s}=2.76$ and 8 TeV.
Furthermore, the statistical uncertainties decreased to $2.7$--$3.2\%$ and $1.4$--$1.8\%$ in the same transverse momentum region.
With the present accuracy, neither for the individual nor for the combined double ratios a significant direct photon excess is observed for $\pt<7$~\GeVc. 

Indeed, in the combined excess ratios an onset of prompt photon production above $\pT>7$ \GeVc\ consistent with expectations from next-to-leading order (NLO) perturbative QCD calculations is visible, but not significant within the given uncertainties.
Three different photon calculations are shown for both systems based on different parton distribution functions and fragmentation functions.
The NLO pQCD calculations~\cite{Jager:2002xm,Paquet:2015lta} are using CT10~\cite{Lai:2010vv,Gao:2013xoa,Guzzi:2011sv} or CTEQ6.1M~\cite{Stump:2003yu} proton PDF and GRV~\cite{Gluck:1992zx} or BFG2~\cite{Bourhis:1997yu} fragmentation functions.
The uncertainty band of the calculation from~\cite{Jager:2002xm} is given by the simultaneous variation of the factorization scale value, $\mu$, ($0.5\pT<\mu<2\pT$) for the factorization, renormalization and fragmentation scales used in the calculation.
Furthermore, a JETPHOX~\cite{Klasen:2017dsy} calculation based on NNPDF2.3QED~\cite{Ball:2013hta} proton PDF and BFG2 FF is provided as well as a POWHEG~\cite{Klasen:2017dsy} calculation based on the same PDF but with the PYTHIA 8 parton shower algorithm instead of a fragmentation function.
The prompt photon expectations in \Fig{fig:DRcombtheory} are calculated as $\Rg^\text{NLO}=1+\frac{\Yd^\text{NLO}}{\Ygd}$ using the particle decay simulation for the contribution of $\Ygd$ to allow a comparison to the double ratios.
With the present uncertainties, the measurements are in agreement with the calculations within their uncertainties over the full measured transverse momentum range.
However, it is not possible to discriminate between the different fragmentation functions and parton distribution functions used in the calculations.

The direct photon spectra, measured as described in \Sect{sec:photonreco} via the multiplication of the inclusive photon spectrum with $1-R_\gamma^{-1}$, are presented in \Fig{fig:incgammadirgamma}.
The horizontal bars at the end of the arrows represent upper limits of direct photon production at 90\% C.L. taking into account the total uncertainty.
Upper limits are calculated for each spectral point where the total uncertainty of the corresponding $\Rg$ point is found to be consistent with unity within $1\sigma$, which is the case for the transverse momenta below 7 \GeVc\ at both collisions energies.
If the value of $\Rg$ including either $1\sigma$ of the statistical or systematic uncertainty is above unity, the direct photon cross section is determined and indicated in the figures with markers and vertical uncertainty bars or boxes in the respective case.
For a direct comparison, the same NLO pQCD calculations as used in \Fig{fig:DRcombtheory} are shown also in \Fig{fig:incgammadirgamma} for the direct photon spectra.
The calculations agree with the measured spectral points within uncertainties and predict a cross section compatible with the determined upper limits at low $\pT$.

\section{Conclusion}
\label{sec:conclusion}
The invariant differential cross sections or upper limits for inclusive and direct photon production in pp collisions at $\sqrt{s}=2.76$ and $8$ TeV were obtained at mid-rapidity and in transverse momentum ranges of $0.4<\pT<10$ \GeVc\ and $0.3<\pT<16$ \GeVc, respectively~(\Fig{fig:incgammadirgamma}).
Photons were reconstructed with the electromagnetic calorimeter and via reconstruction of e$^+$e$^-$ pairs from conversions in the ALICE detector material using the central tracking system, and were combined in the overlapping $\pT$ interval of both methods~(\Fig{fig:incgammaratios}).
Direct photon spectra, or their upper limits at 90\% C.L.\ were extracted using the direct photon excess ratio $\Rg$, which quantifies the ratio of inclusive photons over decay photons generated with a decay-photon simulation~(\Fig{fig:inputspectraandcocktailgammas}).
An additional hybrid method, combining photons reconstructed from conversions with those identified in the EMCal, was included for the combination of the direct photon excess ratio $\Rg$, as well as the extraction of direct photon spectra or their upper limits.
The weighted combination of three consistent measurements~(PCM, \PCMEMC, \EMC) was used to obtain the final direct photon results~(\Fig{fig:DRindmeas}).
At both center-of-mass energies, no significant direct photon signal could be extracted in the explored transverse momentum ranges~(\Fig{fig:DRcombtheory}). 
However, $\Rg$ for $\pT>7$ \GeVc\ is found to be at least one $\sigma$ above unity and consistent with expectations from next-to-leading order pQCD calculations.
Below $7$~\GeVc, total uncertainties of $5$--$12$\% at $\sqrt{s}=2.76$ TeV and $4$--$10$\% at $\sqrt{s}=8$ TeV were achieved.
For this region, upper limits of direct photon production at 90\% C.L. are provided.
Our data limit a possibly enhanced direct photon production at low transverse momentum, and provide a baseline for the interpretation of the direct photon excess observed in heavy-ion collisions.

\newenvironment{acknowledgement}{\relax}{\relax}
\begin{acknowledgement}
\section*{Acknowledgments}
We thank Werner Vogelsang and Jean-Francois Paquet for providing the NLO calculations.

\ifdraft
\else

The ALICE Collaboration would like to thank all its engineers and technicians for their invaluable contributions to the construction of the experiment and the CERN accelerator teams for the outstanding performance of the LHC complex.
The ALICE Collaboration gratefully acknowledges the resources and support provided by all Grid centres and the Worldwide LHC Computing Grid (WLCG) collaboration.
The ALICE Collaboration acknowledges the following funding agencies for their support in building and running the ALICE detector:
A. I. Alikhanyan National Science Laboratory (Yerevan Physics Institute) Foundation (ANSL), State Committee of Science and World Federation of Scientists (WFS), Armenia;
Austrian Academy of Sciences and Nationalstiftung f\"{u}r Forschung, Technologie und Entwicklung, Austria;
Ministry of Communications and High Technologies, National Nuclear Research Center, Azerbaijan;
Conselho Nacional de Desenvolvimento Cient\'{\i}fico e Tecnol\'{o}gico (CNPq), Universidade Federal do Rio Grande do Sul (UFRGS), Financiadora de Estudos e Projetos (Finep) and Funda\c{c}\~{a}o de Amparo \`{a} Pesquisa do Estado de S\~{a}o Paulo (FAPESP), Brazil;
Ministry of Science \& Technology of China (MSTC), National Natural Science Foundation of China (NSFC) and Ministry of Education of China (MOEC) , China;
Ministry of Science, Education and Sport and Croatian Science Foundation, Croatia;
Ministry of Education, Youth and Sports of the Czech Republic, Czech Republic;
The Danish Council for Independent Research | Natural Sciences, the Carlsberg Foundation and Danish National Research Foundation (DNRF), Denmark;
Helsinki Institute of Physics (HIP), Finland;
Commissariat \`{a} l'Energie Atomique (CEA) and Institut National de Physique Nucl\'{e}aire et de Physique des Particules (IN2P3) and Centre National de la Recherche Scientifique (CNRS), France;
Bundesministerium f\"{u}r Bildung, Wissenschaft, Forschung und Technologie (BMBF) and GSI Helmholtzzentrum f\"{u}r Schwerionenforschung GmbH, Germany;
General Secretariat for Research and Technology, Ministry of Education, Research and Religions, Greece;
National Research, Development and Innovation Office, Hungary;
Department of Atomic Energy Government of India (DAE), Department of Science and Technology, Government of India (DST), University Grants Commission, Government of India (UGC) and Council of Scientific and Industrial Research (CSIR), India;
Indonesian Institute of Science, Indonesia;
Centro Fermi - Museo Storico della Fisica e Centro Studi e Ricerche Enrico Fermi and Istituto Nazionale di Fisica Nucleare (INFN), Italy;
Institute for Innovative Science and Technology , Nagasaki Institute of Applied Science (IIST), Japan Society for the Promotion of Science (JSPS) KAKENHI and Japanese Ministry of Education, Culture, Sports, Science and Technology (MEXT), Japan;
Consejo Nacional de Ciencia (CONACYT) y Tecnolog\'{i}a, through Fondo de Cooperaci\'{o}n Internacional en Ciencia y Tecnolog\'{i}a (FONCICYT) and Direcci\'{o}n General de Asuntos del Personal Academico (DGAPA), Mexico;
Nederlandse Organisatie voor Wetenschappelijk Onderzoek (NWO), Netherlands;
The Research Council of Norway, Norway;
Commission on Science and Technology for Sustainable Development in the South (COMSATS), Pakistan;
Pontificia Universidad Cat\'{o}lica del Per\'{u}, Peru;
Ministry of Science and Higher Education and National Science Centre, Poland;
Korea Institute of Science and Technology Information and National Research Foundation of Korea (NRF), Republic of Korea;
Ministry of Education and Scientific Research, Institute of Atomic Physics and Romanian National Agency for Science, Technology and Innovation, Romania;
Joint Institute for Nuclear Research (JINR), Ministry of Education and Science of the Russian Federation and National Research Centre Kurchatov Institute, Russia;
Ministry of Education, Science, Research and Sport of the Slovak Republic, Slovakia;
National Research Foundation of South Africa, South Africa;
Centro de Aplicaciones Tecnol\'{o}gicas y Desarrollo Nuclear (CEADEN), Cubaenerg\'{\i}a, Cuba and Centro de Investigaciones Energ\'{e}ticas, Medioambientales y Tecnol\'{o}gicas (CIEMAT), Spain;
Swedish Research Council (VR) and Knut \& Alice Wallenberg Foundation (KAW), Sweden;
European Organization for Nuclear Research, Switzerland;
National Science and Technology Development Agency (NSDTA), Suranaree University of Technology (SUT) and Office of the Higher Education Commission under NRU project of Thailand, Thailand;
Turkish Atomic Energy Agency (TAEK), Turkey;
National Academy of  Sciences of Ukraine, Ukraine;
Science and Technology Facilities Council (STFC), United Kingdom;
National Science Foundation of the United States of America (NSF) and United States Department of Energy, Office of Nuclear Physics (DOE NP), United States of America.
\fi
\end{acknowledgement}

\bibliographystyle{utphys}
\bibliography{biblio}{}
\newpage
\appendix
\section{The ALICE Collaboration}
\label{app:collab}
\ifdraft
\else

\begingroup
\small
\begin{flushleft}
S.~Acharya\Irefn{org138}\And 
F.T.-.~Acosta\Irefn{org22}\And 
D.~Adamov\'{a}\Irefn{org93}\And 
J.~Adolfsson\Irefn{org80}\And 
M.M.~Aggarwal\Irefn{org97}\And 
G.~Aglieri Rinella\Irefn{org36}\And 
M.~Agnello\Irefn{org33}\And 
N.~Agrawal\Irefn{org48}\And 
Z.~Ahammed\Irefn{org138}\And 
S.U.~Ahn\Irefn{org76}\And 
S.~Aiola\Irefn{org143}
% ...
\And
\renewcommand\labelenumi{\textsuperscript{\theenumi}~}

\section*{Affiliation notes}
\renewcommand\theenumi{\roman{enumi}}
\begin{Authlist}
\item \Adef{orgI}Dipartimento DET del Politecnico di Torino, Turin, Italy
\item \Adef{orgII}M.V. Lomonosov Moscow State University, D.V. Skobeltsyn Institute of Nuclear, Physics, Moscow, Russia
\item \Adef{orgIII}Department of Applied Physics, Aligarh Muslim University, Aligarh, India
\item \Adef{orgIV}Institute of Theoretical Physics, University of Wroclaw, Poland
\end{Authlist}

\section*{Collaboration Institutes}
\renewcommand\theenumi{\arabic{enumi}~}
\begin{Authlist}
\item \Idef{org1}A.I. Alikhanyan National Science Laboratory (Yerevan Physics Institute) Foundation, Yerevan, Armenia
\item \Idef{org2}Benem\'{e}rita Universidad Aut\'{o}noma de Puebla, Puebla, Mexico
\item \Idef{org3}Bogolyubov Institute for Theoretical Physics, National Academy of Sciences of Ukraine, Kiev, Ukraine
\item \Idef{org4}Bose Institute, Department of Physics  and Centre for Astroparticle Physics and Space Science (CAPSS), Kolkata, India
\item \Idef{org5}Budker Institute for Nuclear Physics, Novosibirsk, Russia
\item \Idef{org6}California Polytechnic State University, San Luis Obispo, California, United States
\item \Idef{org7}Central China Normal University, Wuhan, China
\item \Idef{org8}Centre de Calcul de l'IN2P3, Villeurbanne, Lyon, France
\item \Idef{org9}Centro de Aplicaciones Tecnol\'{o}gicas y Desarrollo Nuclear (CEADEN), Havana, Cuba
\item \Idef{org10}Centro de Investigaci\'{o}n y de Estudios Avanzados (CINVESTAV), Mexico City and M\'{e}rida, Mexico
\item \Idef{org11}Centro Fermi - Museo Storico della Fisica e Centro Studi e Ricerche ``Enrico Fermi', Rome, Italy
\item \Idef{org12}Chicago State University, Chicago, Illinois, United States
\item \Idef{org13}China Institute of Atomic Energy, Beijing, China
\item \Idef{org14}Chonbuk National University, Jeonju, Republic of Korea
\item \Idef{org15}Comenius University Bratislava, Faculty of Mathematics, Physics and Informatics, Bratislava, Slovakia
\item \Idef{org16}COMSATS Institute of Information Technology (CIIT), Islamabad, Pakistan
\item \Idef{org17}Creighton University, Omaha, Nebraska, United States
\item \Idef{org18}Department of Physics, Aligarh Muslim University, Aligarh, India
\item \Idef{org19}Department of Physics, Ohio State University, Columbus, Ohio, United States
\item \Idef{org20}Department of Physics, Pusan National University, Pusan, Republic of Korea
\item \Idef{org21}Department of Physics, Sejong University, Seoul, Republic of Korea
\item \Idef{org22}Department of Physics, University of California, Berkeley, California, United States
\item \Idef{org23}Department of Physics, University of Oslo, Oslo, Norway
\item \Idef{org24}Department of Physics and Technology, University of Bergen, Bergen, Norway
\item \Idef{org25}Dipartimento di Fisica dell'Universit\`{a} 'La Sapienza' and Sezione INFN, Rome, Italy
\item \Idef{org26}Dipartimento di Fisica dell'Universit\`{a} and Sezione INFN, Cagliari, Italy
\item \Idef{org27}Dipartimento di Fisica dell'Universit\`{a} and Sezione INFN, Trieste, Italy
\item \Idef{org28}Dipartimento di Fisica dell'Universit\`{a} and Sezione INFN, Turin, Italy
\item \Idef{org29}Dipartimento di Fisica e Astronomia dell'Universit\`{a} and Sezione INFN, Bologna, Italy
\item \Idef{org30}Dipartimento di Fisica e Astronomia dell'Universit\`{a} and Sezione INFN, Catania, Italy
\item \Idef{org31}Dipartimento di Fisica e Astronomia dell'Universit\`{a} and Sezione INFN, Padova, Italy
\item \Idef{org32}Dipartimento di Fisica `E.R.~Caianiello' dell'Universit\`{a} and Gruppo Collegato INFN, Salerno, Italy
\item \Idef{org33}Dipartimento DISAT del Politecnico and Sezione INFN, Turin, Italy
\item \Idef{org34}Dipartimento di Scienze e Innovazione Tecnologica dell'Universit\`{a} del Piemonte Orientale and INFN Sezione di Torino, Alessandria, Italy
\item \Idef{org35}Dipartimento Interateneo di Fisica `M.~Merlin' and Sezione INFN, Bari, Italy
\item \Idef{org36}European Organization for Nuclear Research (CERN), Geneva, Switzerland
\item \Idef{org37}Faculty of Engineering and Science, Western Norway University of Applied Sciences, Bergen, Norway
\item \Idef{org38}Faculty of Nuclear Sciences and Physical Engineering, Czech Technical University in Prague, Prague, Czech Republic
\item \Idef{org39}Faculty of Science, P.J.~\v{S}af\'{a}rik University, Ko\v{s}ice, Slovakia
\item \Idef{org40}Frankfurt Institute for Advanced Studies, Johann Wolfgang Goethe-Universit\"{a}t Frankfurt, Frankfurt, Germany
\item \Idef{org41}Gangneung-Wonju National University, Gangneung, Republic of Korea
\item \Idef{org42}Gauhati University, Department of Physics, Guwahati, India
\item \Idef{org43}Helmholtz-Institut f\"{u}r Strahlen- und Kernphysik, Rheinische Friedrich-Wilhelms-Universit\"{a}t Bonn, Bonn, Germany
\item \Idef{org44}Helsinki Institute of Physics (HIP), Helsinki, Finland
\item \Idef{org45}Hiroshima University, Hiroshima, Japan
\item \Idef{org46}Hochschule Worms, Zentrum  f\"{u}r Technologietransfer und Telekommunikation (ZTT), Worms, Germany
\item \Idef{org47}Horia Hulubei National Institute of Physics and Nuclear Engineering, Bucharest, Romania
\item \Idef{org48}Indian Institute of Technology Bombay (IIT), Mumbai, India
\item \Idef{org49}Indian Institute of Technology Indore, Indore, India
\item \Idef{org50}Indonesian Institute of Sciences, Jakarta, Indonesia
\item \Idef{org51}INFN, Laboratori Nazionali di Frascati, Frascati, Italy
\item \Idef{org52}INFN, Sezione di Bari, Bari, Italy
\item \Idef{org53}INFN, Sezione di Bologna, Bologna, Italy
\item \Idef{org54}INFN, Sezione di Cagliari, Cagliari, Italy
\item \Idef{org55}INFN, Sezione di Catania, Catania, Italy
\item \Idef{org56}INFN, Sezione di Padova, Padova, Italy
\item \Idef{org57}INFN, Sezione di Roma, Rome, Italy
\item \Idef{org58}INFN, Sezione di Torino, Turin, Italy
\item \Idef{org59}INFN, Sezione di Trieste, Trieste, Italy
\item \Idef{org60}Inha University, Incheon, Republic of Korea
\item \Idef{org61}Institut de Physique Nucl\'{e}aire d'Orsay (IPNO), Institut National de Physique Nucl\'{e}aire et de Physique des Particules (IN2P3/CNRS), Universit\'{e} de Paris-Sud, Universit\'{e} Paris-Saclay, Orsay, France
\item \Idef{org62}Institute for Nuclear Research, Academy of Sciences, Moscow, Russia
\item \Idef{org63}Institute for Subatomic Physics, Utrecht University/Nikhef, Utrecht, Netherlands
\item \Idef{org64}Institute for Theoretical and Experimental Physics, Moscow, Russia
\item \Idef{org65}Institute of Experimental Physics, Slovak Academy of Sciences, Ko\v{s}ice, Slovakia
\item \Idef{org66}Institute of Physics, Bhubaneswar, India
\item \Idef{org67}Institute of Physics of the Czech Academy of Sciences, Prague, Czech Republic
\item \Idef{org68}Institute of Space Science (ISS), Bucharest, Romania
\item \Idef{org69}Institut f\"{u}r Kernphysik, Johann Wolfgang Goethe-Universit\"{a}t Frankfurt, Frankfurt, Germany
\item \Idef{org70}Instituto de Ciencias Nucleares, Universidad Nacional Aut\'{o}noma de M\'{e}xico, Mexico City, Mexico
\item \Idef{org71}Instituto de F\'{i}sica, Universidade Federal do Rio Grande do Sul (UFRGS), Porto Alegre, Brazil
\item \Idef{org72}Instituto de F\'{\i}sica, Universidad Nacional Aut\'{o}noma de M\'{e}xico, Mexico City, Mexico
\item \Idef{org73}iThemba LABS, National Research Foundation, Somerset West, South Africa
\item \Idef{org74}Johann-Wolfgang-Goethe Universit\"{a}t Frankfurt Institut f\"{u}r Informatik, Fachbereich Informatik und Mathematik, Frankfurt, Germany
\item \Idef{org75}Joint Institute for Nuclear Research (JINR), Dubna, Russia
\item \Idef{org76}Korea Institute of Science and Technology Information, Daejeon, Republic of Korea
\item \Idef{org77}KTO Karatay University, Konya, Turkey
\item \Idef{org78}Laboratoire de Physique Subatomique et de Cosmologie, Universit\'{e} Grenoble-Alpes, CNRS-IN2P3, Grenoble, France
\item \Idef{org79}Lawrence Berkeley National Laboratory, Berkeley, California, United States
\item \Idef{org80}Lund University Department of Physics, Division of Particle Physics, Lund, Sweden
\item \Idef{org81}Nagasaki Institute of Applied Science, Nagasaki, Japan
\item \Idef{org82}Nara Women{'}s University (NWU), Nara, Japan
\item \Idef{org83}National and Kapodistrian University of Athens, School of Science, Department of Physics , Athens, Greece
\item \Idef{org84}National Centre for Nuclear Research, Warsaw, Poland
\item \Idef{org85}National Institute of Science Education and Research, HBNI, Jatni, India
\item \Idef{org86}National Nuclear Research Center, Baku, Azerbaijan
\item \Idef{org87}National Research Centre Kurchatov Institute, Moscow, Russia
\item \Idef{org88}Niels Bohr Institute, University of Copenhagen, Copenhagen, Denmark
\item \Idef{org89}Nikhef, National institute for subatomic physics, Amsterdam, Netherlands
\item \Idef{org90}NRC ¿Kurchatov Institute¿ ¿ IHEP , Protvino, Russia
\item \Idef{org91}NRNU Moscow Engineering Physics Institute, Moscow, Russia
\item \Idef{org92}Nuclear Physics Group, STFC Daresbury Laboratory, Daresbury, United Kingdom
\item \Idef{org93}Nuclear Physics Institute of the Czech Academy of Sciences, \v{R}e\v{z} u Prahy, Czech Republic
\item \Idef{org94}Oak Ridge National Laboratory, Oak Ridge, Tennessee, United States
\item \Idef{org95}Petersburg Nuclear Physics Institute, Gatchina, Russia
\item \Idef{org96}Physics department, Faculty of science, University of Zagreb, Zagreb, Croatia
\item \Idef{org97}Physics Department, Panjab University, Chandigarh, India
\item \Idef{org98}Physics Department, University of Jammu, Jammu, India
\item \Idef{org99}Physics Department, University of Rajasthan, Jaipur, India
\item \Idef{org100}Physikalisches Institut, Eberhard-Karls-Universit\"{a}t T\"{u}bingen, T\"{u}bingen, Germany
\item \Idef{org101}Physikalisches Institut, Ruprecht-Karls-Universit\"{a}t Heidelberg, Heidelberg, Germany
\item \Idef{org102}Physik Department, Technische Universit\"{a}t M\"{u}nchen, Munich, Germany
\item \Idef{org103}Research Division and ExtreMe Matter Institute EMMI, GSI Helmholtzzentrum f\"ur Schwerionenforschung GmbH, Darmstadt, Germany
\item \Idef{org104}Rudjer Bo\v{s}kovi\'{c} Institute, Zagreb, Croatia
\item \Idef{org105}Russian Federal Nuclear Center (VNIIEF), Sarov, Russia
\item \Idef{org106}Saha Institute of Nuclear Physics, Kolkata, India
\item \Idef{org107}School of Physics and Astronomy, University of Birmingham, Birmingham, United Kingdom
\item \Idef{org108}Secci\'{o}n F\'{\i}sica, Departamento de Ciencias, Pontificia Universidad Cat\'{o}lica del Per\'{u}, Lima, Peru
\item \Idef{org109}Shanghai Institute of Applied Physics, Shanghai, China
\item \Idef{org110}Stefan Meyer Institut f\"{u}r Subatomare Physik (SMI), Vienna, Austria
\item \Idef{org111}SUBATECH, IMT Atlantique, Universit\'{e} de Nantes, CNRS-IN2P3, Nantes, France
\item \Idef{org112}Suranaree University of Technology, Nakhon Ratchasima, Thailand
\item \Idef{org113}Technical University of Ko\v{s}ice, Ko\v{s}ice, Slovakia
\item \Idef{org114}Technische Universit\"{a}t M\"{u}nchen, Excellence Cluster 'Universe', Munich, Germany
\item \Idef{org115}The Henryk Niewodniczanski Institute of Nuclear Physics, Polish Academy of Sciences, Cracow, Poland
\item \Idef{org116}The University of Texas at Austin, Austin, Texas, United States
\item \Idef{org117}Universidad Aut\'{o}noma de Sinaloa, Culiac\'{a}n, Mexico
\item \Idef{org118}Universidade de S\~{a}o Paulo (USP), S\~{a}o Paulo, Brazil
\item \Idef{org119}Universidade Estadual de Campinas (UNICAMP), Campinas, Brazil
\item \Idef{org120}Universidade Federal do ABC, Santo Andre, Brazil
\item \Idef{org121}University College of Southeast Norway, Tonsberg, Norway
\item \Idef{org122}University of Cape Town, Cape Town, South Africa
\item \Idef{org123}University of Houston, Houston, Texas, United States
\item \Idef{org124}University of Jyv\"{a}skyl\"{a}, Jyv\"{a}skyl\"{a}, Finland
\item \Idef{org125}University of Liverpool, Department of Physics Oliver Lodge Laboratory , Liverpool, United Kingdom
\item \Idef{org126}University of Split, Faculty of Electrical Engineering, Mechanical Engineering and Naval Architecture, Split, Croatia
\item \Idef{org127}University of Tennessee, Knoxville, Tennessee, United States
\item \Idef{org128}University of the Witwatersrand, Johannesburg, South Africa
\item \Idef{org129}University of Tokyo, Tokyo, Japan
\item \Idef{org130}University of Tsukuba, Tsukuba, Japan
\item \Idef{org131}Universit\'{e} Clermont Auvergne, CNRS/IN2P3, LPC, Clermont-Ferrand, France
\item \Idef{org132}Universit\'{e} de Lyon, Universit\'{e} Lyon 1, CNRS/IN2P3, IPN-Lyon, Villeurbanne, Lyon, France
\item \Idef{org133}Universit\'{e} de Strasbourg, CNRS, IPHC UMR 7178, F-67000 Strasbourg, France, Strasbourg, France
\item \Idef{org134} Universit\'{e} Paris-Saclay Centre d¿\'Etudes de Saclay (CEA), IRFU, Department de Physique Nucl\'{e}aire (DPhN), Saclay, France
\item \Idef{org135}Universit\`{a} degli Studi di Pavia, Pavia, Italy
\item \Idef{org136}Universit\`{a} di Brescia, Brescia, Italy
\item \Idef{org137}V.~Fock Institute for Physics, St. Petersburg State University, St. Petersburg, Russia
\item \Idef{org138}Variable Energy Cyclotron Centre, Kolkata, India
\item \Idef{org139}Warsaw University of Technology, Warsaw, Poland
\item \Idef{org140}Wayne State University, Detroit, Michigan, United States
\item \Idef{org141}Westf\"{a}lische Wilhelms-Universit\"{a}t M\"{u}nster, Institut f\"{u}r Kernphysik, M\"{u}nster, Germany
\item \Idef{org142}Wigner Research Centre for Physics, Hungarian Academy of Sciences, Budapest, Hungary
\item \Idef{org143}Yale University, New Haven, Connecticut, United States
\item \Idef{org144}Yonsei University, Seoul, Republic of Korea
\end{Authlist}
\endgroup

\fi
\end{document}